\documentclass[11pt]{article}
\usepackage{apacite}
\usepackage{amsfonts}
\usepackage{amsmath}
\usepackage{amssymb}
\usepackage{graphicx}
\usepackage{xcolor, soul}
\usepackage{enumerate}
\usepackage{verbatim}
\usepackage{hyperref}
\usepackage{comment}
\usepackage{url} 
\usepackage{float}
\usepackage{algorithm}
\usepackage{xr}
\usepackage{setspace}
\renewenvironment{abstract}
 {\par\noindent\textbf{\abstractname.}\ \ignorespaces}
 {\par\medskip}

\usepackage[tmargin=1in,bmargin=1in,lmargin=1in,rmargin=1in]{geometry}

\usepackage{sectsty} 
\sectionfont{\fontsize{11}{11}\selectfont}
\subsectionfont{\fontsize{11}{11}\selectfont}

\usepackage[normalem]{ulem}
\usepackage{natbib}

\newcommand{\EXPT}{\mathrm{E}}
\def\bSig\mathbf{\Sigma}

\newcommand{\N}{\mathcal{N}}
\newcommand{\R}{\mathbb{R}}

\newtheorem{theorem}{Theorem}

\newtheorem{condition}{Condition}
\newtheorem{proposition}{Proposition}
\newtheorem{lemma}{Lemma}
\newtheorem{remark}{Remark}

\begin{document}

\def\spacingset#1{\renewcommand{\baselinestretch}%
    {#1}\small\normalsize} \spacingset{1}


{
    \title{Rank-adaptive covariance testing with applications to\\ genomics and neuroimaging}
    \author{David Veitch$^{1}$, Yinqiu He$^{2,*}$, and Jun Young Park$^{1,3,*}$}
    \date{%
    $^1${\small \textit{Department of Statistical Sciences, University of Toronto, Toronto, ON, Canada}}\\
    $^2${\small \textit{Department of Statistics, University of Wisconsin - Madison, 
    Madison, Wisconsin, U.S.A.}}\\
    $^3${\small \textit{Department of Psychology, University of Toronto, Toronto, ON, Canada}}\\
    $^*${\small These authors contributed equally.}\\
}

    \maketitle

\vspace{20mm}

\begin{abstract}
In biomedical studies, testing for differences in covariance may offer scientific insights, especially when differences are driven by complex joint behavior between features. However, when differences in joint behavior are weakly dispersed across many dimensions and arise from differences in low-rank structures within the data, as is often the case in genomics and neuroimaging, existing two-sample covariance testing methods may suffer from power loss. The Ky-Fan($k$) norm, defined by the sum of the top $k$ singular values, is a simple and intuitive matrix norm able to capture signals caused by differences in low-rank structures between matrices, but its statistical properties in hypothesis testing have not been studied well. In this paper, we investigate the behavior of the Ky-Fan($k$) norm in two-sample covariance testing. Ultimately, we propose a novel methodology, Rank-Adaptive Covariance Testing (RACT), which is able to leverage differences in low-rank structures found in the covariance matrices of two groups in order to maximize power. RACT uses permutation for statistical inference, ensuring an exact Type I error control. We validate RACT in simulation studies and evaluate its performance when testing for differences in gene expression networks between two types of lung cancer, as well as testing for covariance heterogeneity in diffusion tensor imaging (DTI) data taken on two different scanner types.
\end{abstract}


\spacingset{1.5}

\vspace{10mm}

\newpage 

\section{Introduction}

\subsection{Covariance testing in biomedical data}

In biomedical studies, 
differences in covariance often offer scientific insights beyond what is inferred by mean differences. In particular, it can help determine if complex joint behavior differs between two groups of samples. In this paper, we present two motivating applications in genomics and neuroimaging that showcase the significance of comparing covariances.

In genomics, gene expression networks, quantified by the covariance of expression levels of multiple genes, provide insights into the genetic drivers of cellular behavior. Important genomic biomarkers can be identified with the help of tests for differences in the gene expression networks between tissue types (e.g., a tumor tissue and normal tissue \citep{park2020integrative}), molecular subtypes of a cancer (e.g., basal and HER2 subtypes in breast cancer), or cancer types (e.g., breast cancer versus ovarian cancer \citep{lock2022bidimensional}). 

In the second motivating example, we consider the `batch effect' problem where non-biological variations are induced by collecting data from multiple sites and scanners. Within neuroimaging, a number of methods have been proposed to estimate and remove these effects \citep{hu2023image}, and have the potential to increase the reliability of scientific findings from the increased sample sizes and more diverse groups of subjects that come from combining datasets. Recently \cite{10.1162/imag_a_00011} and \cite{zhang2024san} have observed batch effects reflected in the heterogeneity of covariances of observations taken at different sites or using different scanners. However, there has been limited work on testing whether these observed covariance heterogeneities are even statistically significant, and hence whether preprocessing the data to mitigate these batch effects is justified.

\subsection{Leveraging low-rank structure}

In both genomics and neuroimaging, techniques leveraging the low-rank structure of the often high dimensional data have been found to be useful in characterizing and understanding variations within the data. These low-rank structures are correspondingly reflected in the spiked structure of the singular values of the covariance matrix. Low-rank structures have been empirically observed in the differences between gene expression networks \citep{amar2013dissection}. In addition, several model-based approaches in cancer genomics demonstrated the utility of leveraging low-rank structures in  studying differences between tumor types \citep{park2020integrative, lock2022bidimensional}. As well, the batch effect induced covariance heterogeneities in neuroimaging appear to be driven by differences in low-rank structures \citep{10.1162/imag_a_00011}. When low-rank structures differ between two groups, it is expected that a low-rank structure will explain most of the differences in covariances between the two groups. Therefore, methods that do not take the inherent low-rank structures in data into account can be underpowered.

\subsection{Literature review}\label{sec:twosamplecovariance}

In both aforementioned application areas, comparing the covariance matrices can be formulated as statistical hypothesis testing problems. In particular, let $\Sigma_1$ and $\Sigma_2$ represent two population covariances from two groups, and the null and alternative hypotheses of interest are given by $H_0:\Sigma_1=\Sigma_2$ and $H_1:\Sigma_1\neq \Sigma_2$. To test $H_0$, utilizing a low-rank structure to characterize the difference in covariance between groups is likely to be useful, however to date few two-sample covariance testing methods have been developed which explicitly do so. 

\cite{schott2007test} and \cite{li2012two} consider test statistics based on the squared Frobenius norm of $\Sigma_1-\Sigma_2$. \cite{srivastava2010testing} develop an estimator based on the trace of $\Sigma_1$ and $\Sigma_2$ as well as $\Sigma_1^2$ and $\Sigma_2^2$. \cite{cai2013two} proposes a test powerful against sparse alternatives based on the maximum of standardized elementwise differences of sample covariance matrices between two groups. \cite{danaher2015covariance} is a biological pathway inspired test which uses the leading eigenvalues and trace of the sample covariance matrices. \cite{zhu2017testing} is based on a sparsity-constrained leading eigenvector of $\Sigma_1-\Sigma_2$. The test statistic proposed by \cite{he2018high}  is based on differences in superdiagonals between $\Sigma_1$ and $\Sigma_2$, which is particularly powerful when $\Sigma_1$ and $\Sigma_2$ have a banded structure. \cite{ding2023sampletestcovariancematrices}, building upon their work in random matrix theory \citep{ding2023global}, define a small neighborhood around the median eigenvalue from the sample covariance matrix from one group, and use this neighborhood to test whether the eigenvalues from the sample covariance matrices of both groups differ. While individually these methods may exhibit high power under the alternative hypothesis for certain low-rank structures, they risk being underpowered for more general forms of low-rank structures.

\subsection{Our contributions}\label{sec:challengesandopportunities}

In this paper, we bridge this gap and propose a two-sample covariance testing method able to adapt to the form of the low-rank structures within the data. Specifically, our test statistic is adaptive to the form of the low-rank structure found in the difference of two sample covariance matrices, and we utilize a permutation scheme to ensure strict Type I error control in the finite-sample setting. 

The remainder of the paper is as follows. In Section \ref{sec:meth} our test statistic is presented, with the asymptotic behavior of the test statistics included in RACT presented in Section \ref{sec:theory}. In Section \ref{sec:simulationstudy} we conduct simulation studies to demonstrate its  adaptivity to various forms of covariance differences and compare its performance to other tests. We then apply RACT to the two application areas of interest in Section \ref{sec:realdataanalysis}. In Section \ref{sec:discussion} we discuss potential extensions and limitations of RACT.

\section{Methodology}
\label{sec:meth}

\subsection{Notation and setup}

Let $X_1^{(1)},\dots,X_{n_1}^{(1)}$, $X_1^{(2)},\dots,X_{n_2}^{(2)} \in\mathbb{R}^p$ denote the observed features from two groups of sample sizes $n_1$ and $n_2$, respectively, with equal population means (for practical purposes as a preprocessing step the data can be centered using the sample mean for each group), and let $(\Sigma_1,\Sigma_2)$ and $(\widehat{\Sigma}_1,\widehat{\Sigma}_2)$ be the population and sample covariance matrices of these groups, respectively. In addition, we define $n=n_1+n_2$ to be the total number of samples and $\widehat{\Sigma}$ to be the sample covariance calculated using all observations from both groups. In this section, we test for differences in the covariance matrix. However, our proposed idea is general and could be similarly extended to correlation matrices. We define our null and alternative hypotheses as $H_0:\Sigma_1=\Sigma_2$ and $H_{1}:\Sigma_1\neq \Sigma_2$.

\subsection{Ky-Fan($k$) statistic - fixed $k$}\label{sec:kyfanfixedk}

Testing $H_0: \Sigma_1 = \Sigma_2$ requires quantifying the difference between $\widehat{\Sigma}_1$ and $\widehat{\Sigma}_2$. As noted in Section \ref{sec:twosamplecovariance}, several existing two-sample covariance testing methods are based on test statistics which utilize a limited number of singular values or the Frobenius norm  for detecting differences in covariance structures. However, these test statistics may only be powerful under certain covariance differences. For example, a test statistic based on a single singular value \citep{zhu2017testing}, would be underpowered when several singular values drive the difference in covariance. On the other hand, methods such as  \cite{schott2007test} and \cite{li2012two} which use the Frobenius norm, given by $||\widehat\Sigma_1-\widehat\Sigma_2||_F=\sqrt{\sum_{r=1}^p\sum_{s=1}^p (\widehat\Sigma_1[r,s]-\widehat\Sigma_2[r,s])^2}$, is well-suited when each entry of  $\widehat{\Sigma}_1-\widehat{\Sigma}_2$ has non-zero expectation (i.e., dense signals), but it could be underpowered when the entries with non-zero expected values are sparse. Also, even when the signals are dense, considering that the Frobenius norm of a matrix is equivalent to the square root of the sum of squares of all singular values of the matrix, it may be underpowered when ${\Sigma}_1-{\Sigma}_2$ is low-rank.

For a fixed $k\in\mathbb{N}$, one way to characterize the difference between sample covariance matrices is through the use of a Ky-Fan($k$) norm defined by
\begin{align}
    T_k &= ||\widehat{\Sigma}_1-\widehat{\Sigma}_2||_{(k)} \label{eqn:Tk},
\end{align}
where $ \| A\|_{(k)} = \sum_{l=1}^k \sigma_{l}(A)$ and $\sigma_{l}(A)$ is the $l$th largest singular value of a matrix $A$. If $\Sigma_1-\Sigma_2$ is low-rank, there may exist $k<p$ such that this Ky-Fan($k$) norm captures most of the variation of the signal, and by excluding the bottom $p-k$ singular values, ignores noise introduced through finite-sample variability. The way in which we characterize the difference in covariance via Ky-Fan($k$) norm statistics which only involve a subset of the singular values from $\widehat{\Sigma}_1-\widehat{\Sigma}_2$ is similar in spirit to recent work by \cite{ding2023sampletestcovariancematrices}. Their testing procedure first estimates the median eigenvalue of the covariance matrix for one group, and then for two groups compares the sum of the eigenvalues which lie close to this median.

\subsection{Adaptive Ky-Fan$(k)$ statistic}\label{sec:kyfan}

Although $T_k$ is a well-motivated test statistic for a prespecified $k$, it is unclear in advance what value of $k$ will maximize power. For small $k$, $T_k$ may fail to capture the signal that exists outside the top-$k$ singular values of $\widehat{\Sigma}_1-\widehat{\Sigma}_2$. Alternatively, if $k$ is chosen too large then some of the singular values included in $T_k$ will be noise, decreasing the signal to noise ratio of $T_k$. Section \ref{sec:theory} further discusses this signal-to-noise trade-off in the asymptotic setting. 

The problem of selecting the `optimal' $k$ motivates the proposed method, rank-adaptive  covariance testing (RACT), which considers  the adaptive test statistic:
\begin{align}
    T_{\text{RACT}} &= \max_{k\in\mathcal{K}} \frac{T_k - \text{E}_{H_0}[T_k]}{\sqrt{\text{Var}_{H_0}[T_k]}},\label{eqn:maxT}
\end{align}
where $\mathcal{K}=\{1,\dots,K\}$ represents the collection of Ky-Fan($k$) norms from 1 to $K$, and $\text{E}_{H_0}[T_k]$, $\text{Var}_{H_0}[T_k]$ are the expected value and variance of $T_k$ under $H_0$.

In Section \ref{sec:theory}, for normal data in the asymptotic setting, we will show that the signal-to-noise ratio of $T_k$ is formulated as a function of  $k$, $\Sigma_1$, and $\Sigma_2$. Therefore, when $\Sigma_1$ and $\Sigma_2$ are unknown it is difficult to ascertain in advance which $k\in\mathbb{N}$ will attain the maximum signal-to-noise ratio. By taking a maximum across different values of $k$ for an appropriately normalized $T_k$, under the alternative, the behavior of $T_\text{RACT}$ will resemble the behavior of the signal-to-noise maximizing $T_k$, and similarly the power of $T_\text{RACT}$ will be close to the power of this $T_k$ (our simulations in Section \ref{sec:simulationstudy} reflect this adaptivity). By including a diverse set of norms in $T_\text{RACT}$, an investigator is freed from having to make a potentially consequential decision as to what norm should be used for testing $H_0$, and at the same time potentially benefiting from the inclusion of norms sensitive to certain types of structures in $\Sigma_1-\Sigma_2$.

The maximum $K$ can be chosen either as $\min(n,p)$, or a smaller value that reflects prior knowledge of the data or computational considerations (since truncated SVD for the top $K$ singular values is an $O(K\times \min(n,p)^2)$ operation). In this paper, we choose $K$ to be the smallest $K\leq \min(n,p)$ such that the variation of $\widehat\Sigma$ explained by its top $K$ singular values exceeds 80\%. This is done as the bottom 20\% of the singular values are unlikely to provide much of the signal for our test statistic in the high-dimensional setting, and by including observations from both groups we ensure $K$ is the same for all permutations of the data.

\subsection{Permutation testing}\label{sec:permtesting}

We use permutation to estimate $\text{E}_{H_0}[T_k]$ and $\text{Var}_{H_0}[T_k]$, as well as to calculate a $p$-value for conducting hypothesis testing based on $T_{\text{RACT}}$. The use of permutation to calculate a $p$-value is attractive due to the dependencies inherent in the $T_k$ statistics.  We create $B$ permuted datasets (randomly permuting the subjects between the two groups) and take the empirical means and standard deviations of $T_k$ for all $k\in\mathcal{K}$ to estimate $\text{E}_{H_0}[T_k]$ and $\text{Var}_{H_0}[T_k]$ ({note since $K$ is calculated using all observations, and is hence the same across all permutations, $\mathcal{K}$ is also the same across permutations}). Then, using these $B$ permuted datasets, we calculate $\{T_\text{RACT}^{(1)},\dots,T_\text{RACT}^{(B)}\}$ where $T_\text{RACT}^{(b)}$ is calculated in the same way as in \eqref{eqn:maxT} except it uses a permuted dataset, and we calculate the $p$-value of our observed statistic as $p_\text{RACT}=\left[1+\sum_{b=1}^B I(T_\text{RACT}^{(b)}\geq T_\text{RACT}) \right]/[B+1]$.

\begin{remark}\label{rem:tractminp} \normalfont

While Section \ref{sec:theory} suggests that for normally distributed data in the asymptotic setting these statistics for certain values of $k$ are marginally normal, making the standardized $T_k$ comparable across different $k$, for finite sample sizes we observe these statistics may deviate from normality for small $k$. We find a minimum $p$-value approach in smaller samples enhances the adaptivity of RACT, where the test statistic is defined by $T_{\text{RACT-min}p}=\min_{k\in\mathcal{K}}p_k$
where $p_k$ is the $p$-value corresponding  to an individual $T_k$ (see Web Appendix A for details).
\end{remark}

\section{Theoretical analysis}\label{sec:theory}

In this section, we provide theoretical understanding towards the proposed statistic and testing procedure under the null and alternative hypotheses, respectively. To facilitate the discussion,  
we consider two-sample independent  observations $X^{(1)}_1,\ldots, X^{(1)}_{n_1}\sim \mathcal{N}_p(0,\Sigma_1)$ and $X^{(2)}_1,\ldots, X^{(2)}_{n_2}\sim  \mathcal{N}_p(0,\Sigma_2)$ throughout this section. 

\subsection{Null hypothesis $\Sigma_1=\Sigma_2$} \label{sec:null_theory_results}

Under the null hypothesis $H_0: \Sigma_1=\Sigma_2$, we show that 
using the proposed statistic along with the permutation procedure can effectively control the Type I error given a finite sample size \citep{lehmann2021testing}, which is formally stated below. 

\begin{proposition}\label{thm:permutationconsit}
Under $H_0$, given any significance level $\alpha\in (0,1)$, the permutation test based on $p_{\operatorname{RACT}}$ in Section \ref{sec:permtesting} has size $\alpha$, i.e.,  
$\mathbb{P}_{H_0}(p_{\operatorname{RACT}}\leqslant \alpha) = \lfloor (B+1)\alpha \rfloor/(B+1) \leqslant \alpha $, where $B$ denotes the number of permutations, and $\lfloor \cdot \rfloor$ represents the floor function.  
\end{proposition} 

The permutation procedure is applicable regardless of the structure of $\Sigma_1=\Sigma_2$ under $H_0$. On the other hand, it is unclear whether there exists a universal distribution that can characterize the asymptotic distribution of $T_{\mathrm{RACT}}$ under $H_0$. Web Appendix C.1 shows the empirical distribution of $T_k$ for various $k$ under $H_0$ across five covariance structures. Based on these empirical results, it appears the properties of these distributions, in particular their skewness and tail behavior, are not uniform, suggesting $T_\text{RACT}$'s distribution may not be uniformly characterized 
across different covariance structures.

\subsection{ Alternative hypotheses $\Sigma_1\neq \Sigma_2$}

Under alternative hypotheses,
we examine asymptotic power of the proposed adaptive statistic $T_{\mathrm{RACT}}$.  For  ease of understanding, we next make extra regularity conditions on the underlying distribution. We emphasize that these conditions are assumed primarily to obtain simple analytical forms below and will not restrict our method's practical use, given we use permutation instead of asymptotic results. 
In the following section, we present asymptotic results, {with} the proofs {relegated to} 
Web Appendix B. 

\begin{condition} \label{cond:regularity}
As $n=n_1+n_2\to \infty$, assume (i) $r_1=n/n_1$ and  $r_2=n/n_2$ remain bounded,  (ii)  $p/\sqrt{n}\to 0$, (iii) 
the singular values of $\Sigma_1$ and $\Sigma_2$ are bounded away from zero and infinity, and 
  (iv)  $\Sigma_1-\Sigma_2$ has finite rank, and its nonzero eigenvalues are bounded away from zero with non-vanishing eigen gaps.
\end{condition}

\begin{theorem} \label{thm:twosamlimitkyfan_1}
Assume the distribution of data satisfies Condition \ref{cond:regularity}. 
Under $H_A:\Sigma_1\neq \Sigma_2$, 
for $1\leqslant k\leqslant \mathrm{rank}(\Sigma_1-\Sigma_2)$,
\begin{align} \label{eq:powerform1_2}
 \frac{\sqrt{n}}{w_{1:k}}\left\{T_k-\|\Sigma_1-\Sigma_2\|_{(k)}\right\}\xrightarrow{d} \mathcal{N}(0,1), 
\end{align}
where $\omega_{1:k}^2=2\sum_{s=1}^2r_s\mathrm{tr}\{(U_k^{\top}\Sigma_s V_k)^2\}$ and $U_k$ and $V_k$ represent the left and right singular vectors of $\Sigma_1-\Sigma_2$ corresponding to the largest $k$ singular values.

 At a fixed significance level $\alpha\in (0,1)$, suppose the test threshold $t$ satisfies $\mathbb{P}_{H_0}(T_{\mathrm{RACT}} \geqslant t) \in [c_0\alpha, \alpha]$ for a constant $c_0\in (0,1)$, i.e., it controls the  Type I error without being overly conservative. Assume finite values of $k$ is used in $\mathcal{K}$ for  $T_{\mathrm{RACT}}$. 
Then as $n\to \infty$, 
\begin{align} \label{eq:consistent_power1}
     \mathbb{P}_{H_A}(T_{\mathrm{RACT}}\geqslant t) \to 1.
\end{align} 
\end{theorem}

Equation \eqref{eq:consistent_power1} suggests that our test based on $T_{\mathrm{RACT}}$ is consistent against $H_A$ under Condition \ref{cond:regularity}. It is derived based on \eqref{eq:powerform1_2}, 
the asymptotic distribution of individual $T_k$, and the fact that
\begin{align} \label{eq:ract_lower_power}
    \mathbb{P}_{H_A}(T_{\mathrm{RACT}}\geqslant t) \geqslant \,  \max_{k\in \mathcal{K}}    \mathbb{P}_{H_A} (T_{k}\geqslant t_k), 
\end{align} 
where $t_k = t\sqrt{\mathrm{Var}_{H_0} [T_k] } + \mathrm{E}_{H_0}[T_k]$. 
Equation \eqref{eq:ract_lower_power} 
indicates  $T_{\mathrm{RACT}}$ can adaptively combine the information from $T_k$ for $k\in \mathcal{K}$.

In addition, the asymptotic normality in 
 \eqref{eq:powerform1_2}
implies  $\mathbb{P}_{H_A}(T_k\geqslant t_k) $, the rejecting probability of $T_k$ at a threshold $t_k$ under $H_A$, is approximately    
\begin{align} \label{eq:phi_approx_power}
     1-\Phi\left[   \frac{\sqrt{n}}{w_{1:k}}\{t_k-\|\Sigma_1-\Sigma_2\|_{(k)} \}\right], 
\end{align}
where $\Phi(\cdot)$ denotes the cumulative distribution function of the standard  normal distribution. 
Our proof of Theorem \ref{thm:twosamlimitkyfan_1} indicates  that $t_k=o\{\|\Sigma_1-\Sigma_2\|_{(k)}\}$ under suitable conditions.
In this case,  \eqref{eq:phi_approx_power} is dominated by the term involving the signal-to-noise ratio $\mathrm{SNR}_k:=\|\Sigma_1-\Sigma_2\|_{(k)}/w_{1:k}$. As $\Sigma_1-\Sigma_2$ varies, a higher value of $\mathrm{SNR}_k$  indicates a larger value of \eqref{eq:phi_approx_power}. To gain insight into how the rejecting probability of $T_k$ varies with respect to $k$  under $H_A$, 
Proposition \ref{prop:interpret} below examines   $\operatorname{SNR}_k$ versus $k$. 

\begin{proposition}\label{prop:interpret}
For two indexes $k_1<k_2 \in \{1,\ldots, p\}$,  $\operatorname{SNR}_{k_2}\geq \operatorname{SNR}_{k_1}$ if and only if $\beta_{k_1,k_2}  \geq \sqrt{\gamma_{k_1,k_2}+1} - 1$, where we define
\begin{align*}
	\beta_{k_1,k_2}=&~\frac{\|\Sigma_1-\Sigma_2\|_{(k_2)}-\|\Sigma_1-\Sigma_2\|_{(k_1)}}{\|\Sigma_1-\Sigma_2\|_{(k_1)}}=\frac{\sum_{j=k_1+1}^{k_2}\lambda_{j}}{\|\Sigma_1-\Sigma_2\|_{(k_1)}}, \\[3pt]
 \gamma_{k_1,k_2}=&~\frac{\omega_{1:k_2}^2-\omega^2_{1:k_1}}{\omega_{1:k_1}^2}
\end{align*}
as the relative increments of signal and noise, respectively. 
\end{proposition}

In Proposition \ref{prop:interpret}, 
$ \beta_{k_1,k_2}\geq \sqrt{\gamma_{k_1,k_2}+1}-1$ can be interpreted as the relative signal increment being larger than the relative variance increment.  
This suggests that even when the signal has a positive increment, whether the test power can be enhanced or not depends on the  trade-off between signal and noise increments. 
For example, consider $\Sigma_1=cI$,  $\Sigma_2=\Sigma_1+\mathrm{diag}(4,1,0,\ldots, 0)$, and balanced two sample $n_1=n_2$. We have $\beta_{1,2} = 1/4$, 
and  $\gamma_{1,2}  =  [ c^2 + (c+1)^2] / [ c^2 + (c+2)^2] $. 
If $c=1$,  $ \beta_{1,2} > \sqrt{\gamma_{1,2}+1}-1$, implying $\mathrm{SNR}_{2} > \mathrm{SNR}_{1}$.
This suggests  that adding the second largest singular value to $T_{\mathrm{RACT}}$ is likely to enhance the test power.  
On the contrary,  if $c=10 $, the reversed inequality $\mathrm{SNR}_{2} < \mathrm{SNR}_{1}$ holds, 
implying that including the second largest singular value in $T_{\mathrm{RACT}}$ may not improve the test power. 
This trade-off shows the difficulty of determining an optimal $k$ exactly in observational data. It partially justifies the proposed adaptive version which integrates multiple Ky-Fan$(k)$ statistics in Equation \eqref{eqn:maxT}.

\section{Simulation study}\label{sec:simulationstudy}

\subsection{Simulation setup}\label{sec:simulationsetup}
We conducted extensive simulation studies to investigate  the performance of RACT. All simulations, unless otherwise noted, were run in the representative setting of $n_1=n_2=25$ and $p=250$. For simulations related to the null hypothesis, the covariance matrix was $\Sigma_1$ for all observations; otherwise $\Sigma_1$ and $\Sigma_2$ represent the covariance matrices for group 1 and group 2 respectively.  We simulated data from covariance matrices with low-rank structures similar to those commonly found in biomedical data.  In the data simulation settings S1-S3 below, we let $\Sigma_1=I+UU^\top$ and $\Sigma_2=I+VV^\top$  where $U$ and $V$ are low-rank matrices, so that $\Sigma_1-\Sigma_2$ is low-rank. We define $w_U,w_V$ as the ranks of $U$ and $V$, and in our simulations either $w_U=w_V=2$ or $w_U=w_V=5$. To randomly generate appropriately sized blocks of the low-rank components of the covariance matrix $U_1U_1^\top,U_2U_2^\top,V_1V_1^\top$, we generated the columns of $U_1,U_2,V_1$ independently using the first $w$ singular vectors from randomly generated matrices $A_1A_1^\top,A_2A_2^\top,A_3A_3^\top$, where $A_1,A_2,A_3$ are appropriately sized matrices with 
independent and identically distributed standard normal entries.
Figure \ref{fig:covariance_structure} is provided for graphical illustrations of our simulation settings.

\begin{figure}[H]
\centering
\includegraphics[width=\textwidth]{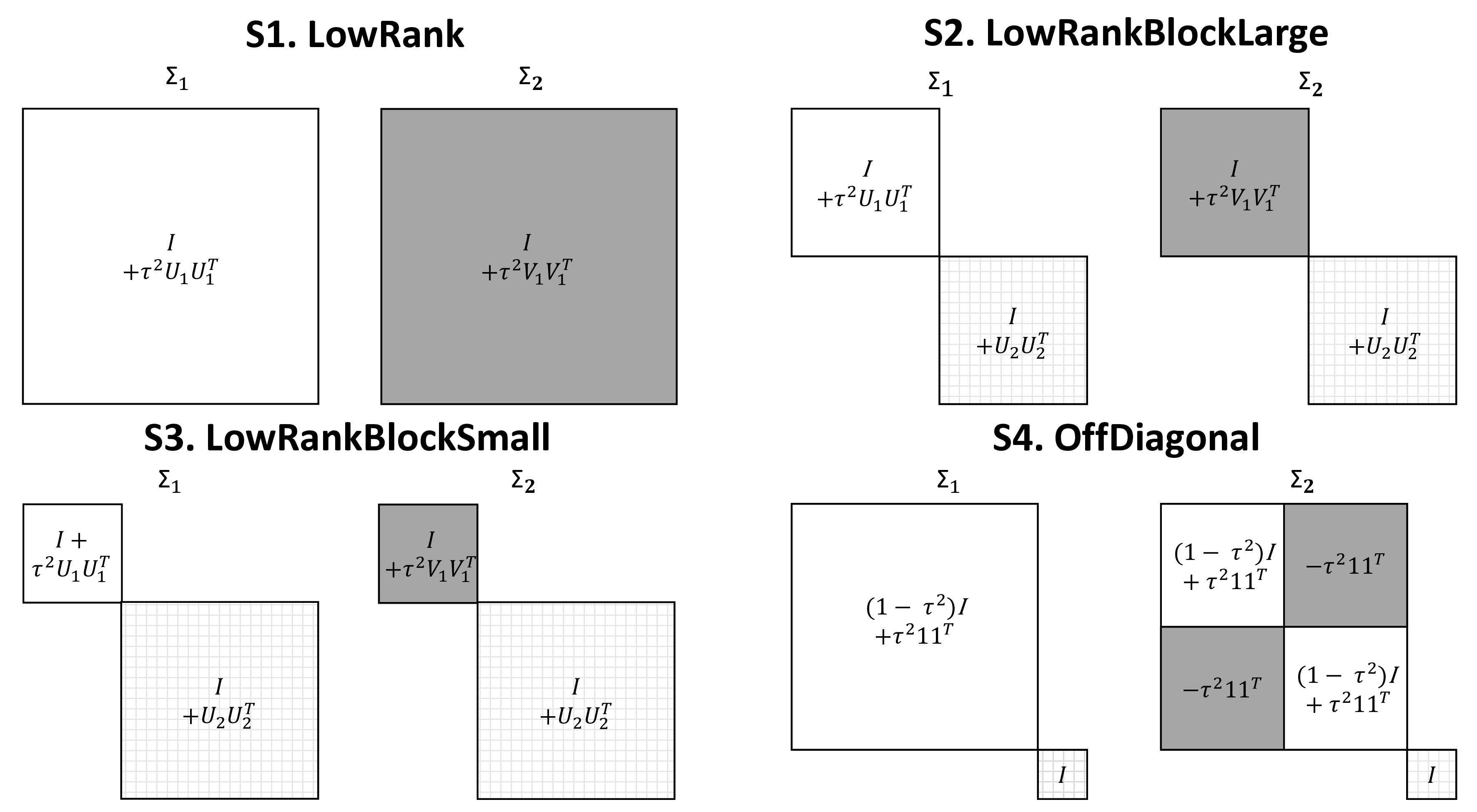}
	\caption{Visualizations of the covariance matrices of both groups across the four simulation scenarios. Low-rank, and low-rank block structures such as these are commonly found in biomedical data.} 
 \label{fig:covariance_structure}
\end{figure}

\begin{enumerate}
    \item[S1.] (LowRank): We set $UU^\top=\tau^2U_1U_1^\top,VV^\top=\tau^2V_1V_1^\top$ with $U_1U_1^\top,V_1V_1^\top\in\mathbb{R}^{p \times p}$. Here, all elements of the covariance matrix experience a change with high probability. 
    \item[S2.] (LowRankBlockLarge): The difference in covariance is localized within one large diagonal block
    \begin{align}
        UU^\top=\begin{bmatrix}\tau^2U_1U_1^\top &0\\0&U_2U_2^\top \end{bmatrix} \;\;\; VV^\top=\begin{bmatrix}\tau^2V_1V_1^\top &0\\0&U_2U_2^\top\end{bmatrix} \;\;\; U_1U_1^\top,V_1V_1^\top,U_2U_2^\top\in\R^{p/2\times p/2}. \nonumber
    \end{align} 
    \item[S3.] (LowRankBlockSmall): The difference in covariance is present in a small block along the diagonal. Similar to S2 except $U_1U_1^\top,V_1V_1^\top \in\R^{10\times 10}$ and $U_2U_2^\top \in \R^{(p-10)\times(p-10)}$. 
    \item[S4.] (OffDiagonal): The difference in covariance is localized within  an off-diagonal block. $\Sigma_1=\begin{bmatrix} A_1 &0\\0&I\end{bmatrix},\Sigma_2=\begin{bmatrix} A_2 &0\\0&I\end{bmatrix}$ where $A_1,A_2,I \in \R^{p/2 \times p/2}$. $A_1$ has an equicorrelation structure in that $a_{r,r}=1$ for $r\in 1,\dots,p/2$ and $a_{r,s}=\tau^2$ for $r\neq s$. $A_2$ is equal to $A_1$ except the covariances between dimensions $(1,\dots,\lceil p/4 \rceil-1)$ and $(\lceil p/4 \rceil,\dots,p/2)$ are set to $-\tau^2$.
    
\end{enumerate}

To evaluate Type I error, we generated 10000 independent datasets and used  2000 permutations for each dataset. For power analysis we used 1000 independent datasets and 1000 permutations. In all simulations we use the $T_\text{RACT-min$p$}$ statistic from Remark \ref{rem:tractminp}.

\subsection{Simulation results}

\subsubsection{Control of Type I error}\label{sec:type1error}

RACT provided reliable Type I error control in all of the simulation setups, which was expected from the permutation scheme. Across S1-S4 for $\alpha=0.05$ RACT's Type I error fell between 0.048 and 0.055. 

\subsubsection{Power comparison of RACT to individual Ky-Fan($k$) norm-based tests ($T_k$)}\label{sec:powersims}

\begin{figure}[h]
	\begin{center}
	\includegraphics[width=0.9 \linewidth]{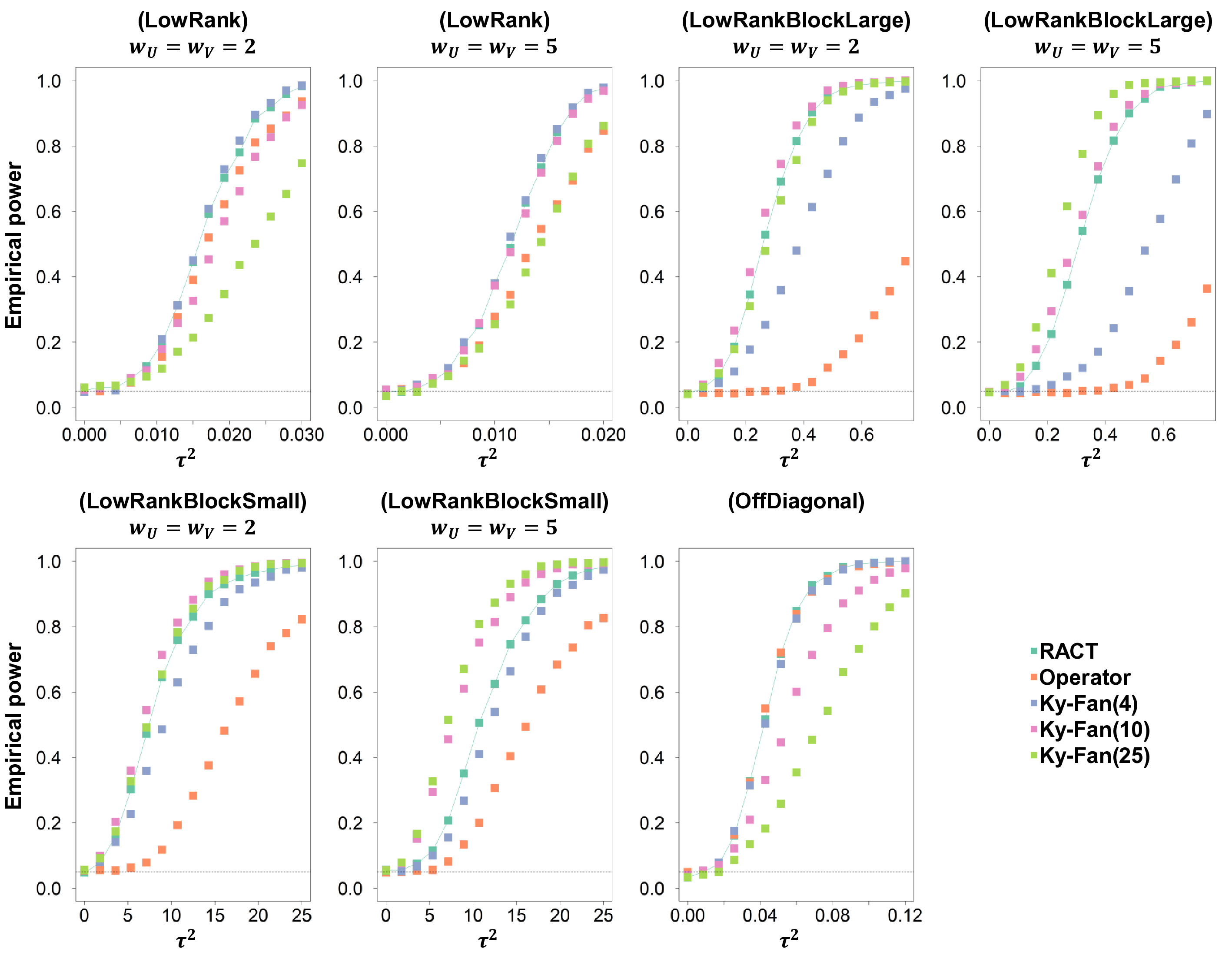}
	\end{center}
	\caption{Empirical power curves when using select single norms, as well as when using RACT method. $n=50,p=250$ for all simulations, and dotted line shows prescribed Type I error rate. Across most simulation settings the power of RACT is close to that of the best performing Ky-Fan($k$) norm. For (LowRankBlockSmall) $w_U=w_V=5$, a large unchanging block leads to a suboptimal choice of $K$, limiting RACT's power.} 
 \label{fig:simulation}
\end{figure}

 Figure \ref{fig:simulation} shows the power of RACT when specific single norms were used, as well as when multiple norms were included as described in Section \ref{sec:kyfan}. The individual norms we present are the operator (i.e., the Ky-Fan(1) norm), Ky-Fan(4), Ky-Fan(10), and Ky-Fan(25). For $w_U=w_V=2$ simulations, the Ky-Fan(4) norm demonstrates the power of a Ky-Fan($k$) norm close to the rank of $\Sigma_1-\Sigma_2$, and the Ky-Fan(10) does so similarly for the $w_U=w_V=5$  setting. The Ky-Fan(1) and Ky-Fan(25) norms are representative norms used to show the power of Ky-Fan($k$) norms for small and large values of $k$ relative to the rank of $\Sigma_1-\Sigma_2$.  Recall that because RACT selects $K$ based on the  covariance matrix using all observations, RACT may not include all of the individual norms we compare it to in all simulations.

\begin{enumerate}
    \item[S1.] (LowRank): For S1 when $w_U=w_V=2$ RACT's power was similar to that of the Ky-Fan(4) norm, which we would expect would have high power as $\Sigma_1-\Sigma_2$ is of rank at most 4. Increasing the rank such that $w_U=w_V=5$  led to an improvement in the performance of the Ky-Fan(10) and Ky-Fan(25) norms and a relative underperformance of the operator norm. Interestingly for the $w_U=w_V=5$ case the Ky-Fan(4) norm still achieved high power, reflecting the challenge of ascertaining in advance the ideal set of Ky-Fan($k$) norms for RACT to include in the finite-sample setting.
    \item [S2.] (LowRankBlockLarge): When a large block experienced a change, the  operator norm markedly underperformed.  On average RACT chose $K$ as 18.8 and 11.4 for the $w_U=w_V=2$ and $w_U=w_V=5$ cases respectively. These choices of $K$ contributed to RACT's high power, since the Ky-Fan($k$) norms for large values of $k$ were particularly powerful.
    \item [S3.] (LowRankBlockSmall): As the change is relegated to a small block we again see the Ky-Fan(10) and Ky-Fan(25) norms exhibit increased power. The  $w_U=w_V=5$ setting demonstrates the way in which the difference occurring in a very small block represents a challenging scenario for RACT. Here RACT's power was lower than the Ky-Fan(10) and Ky-Fan(25) norms  as it suboptimally chose $K$ due to the large singular values associated with the unchanging block. Across all simulations RACT's average choice of $K$ was 9.2 and 7.2 for the $w_U=w_V=2$ and $w_U=w_V=5$ cases respectively, implying that the more powerful Ky-Fan($k$) norms (i.e., those with larger values of $k$) were often excluded from RACT's adaptive test statistic. 
    \item [S4.] (OffDiagonal): This setting sees two very large singular values in $\Sigma_1-\Sigma_2$, and this is reflected in the strong performance of the operator norm. Notable was the poor performance of the Ky-Fan($25$) norm, an indication that including unsuitable Ky-Fan($k$) norms in RACT may not contribute to improved power.
\end{enumerate}

A notable result seen in S2 and S3, but not in S1, is the strong performance of the Ky-Fan(25) norm, despite the fact that the rank of $\Sigma_1-\Sigma_2$ is  at most 4 and 10 for  $w_U=w_V=2$ and $w_U=w_V=5$ respectively. This can be attributed to the fact that for S2 and S3 in the finite-sample setting the top singular values of $\widehat{\Sigma}_1-\widehat{\Sigma}_2$ can be driven by the unchanging block of $\Sigma_1$ and $\Sigma_2$ for small values of $\tau^2$. This led to the signal arising from the changing block to be excluded from the Ky-Fan($k$) norms for small values of $k$, reducing these norms' power. This points to the difficulty of ascertaining the optimal $K$ in advance as the interaction of the covariance structures, $n$, and $\tau^2$, seem to determine which Ky-Fan($k$) norms produce the highest power. Relatedly, in S2 and S3 RACT performs similarly to the best performing norm in the $w_U=w_V=2$ setting, but not in the $w_U=w_V=5$ setting. In both cases this appears to be caused by $K$ being chosen relatively lower in the $w_U=w_V=5$ setting, and hence the higher power norms were excluded from $T_\text{RACT}$, decreasing its power. 

\subsubsection{Comparison of RACT with other methods}\label{sec:sims_compare}

We compare RACT to four other methods, each denoted by their authors' initials: SY \citep{srivastava2010testing}, LC \citep{li2012two}, CLX \citep{cai2013two}, and HC \citep{he2018high} with the implementations found in the \texttt{R} package \texttt{UHDtst} (\url{https://github.com/xcding1212/UHDtst}). Since all of the competing methods considered are based on asymptotic results that may report inflated Type I error rates in small sample sizes, for a fair comparison in our power simulations we implement a permutation-based version of all of these methods (i.e., we use the implementation found in \texttt{UHDtst} and then use permutation to find a critical value with controlled Type I errror). For the hyperparameters used for each method, we follow the default implementation in \texttt{UHDtst}; namely HC tests $\lfloor p^{0.7}\rfloor$ superdiagonals. We note that the \texttt{UHDtst} package also includes the two-sample covariance testing method of \cite{ding2023sampletestcovariancematrices}, however  given its unreliable Type I error control in its original implementation for the sample sizes we consider, and the computational cost of modifying their method to implement a permutation-based version, we do not include it in our analysis.

\begin{figure}[H]
\centering
\includegraphics[width=0.9 \linewidth]{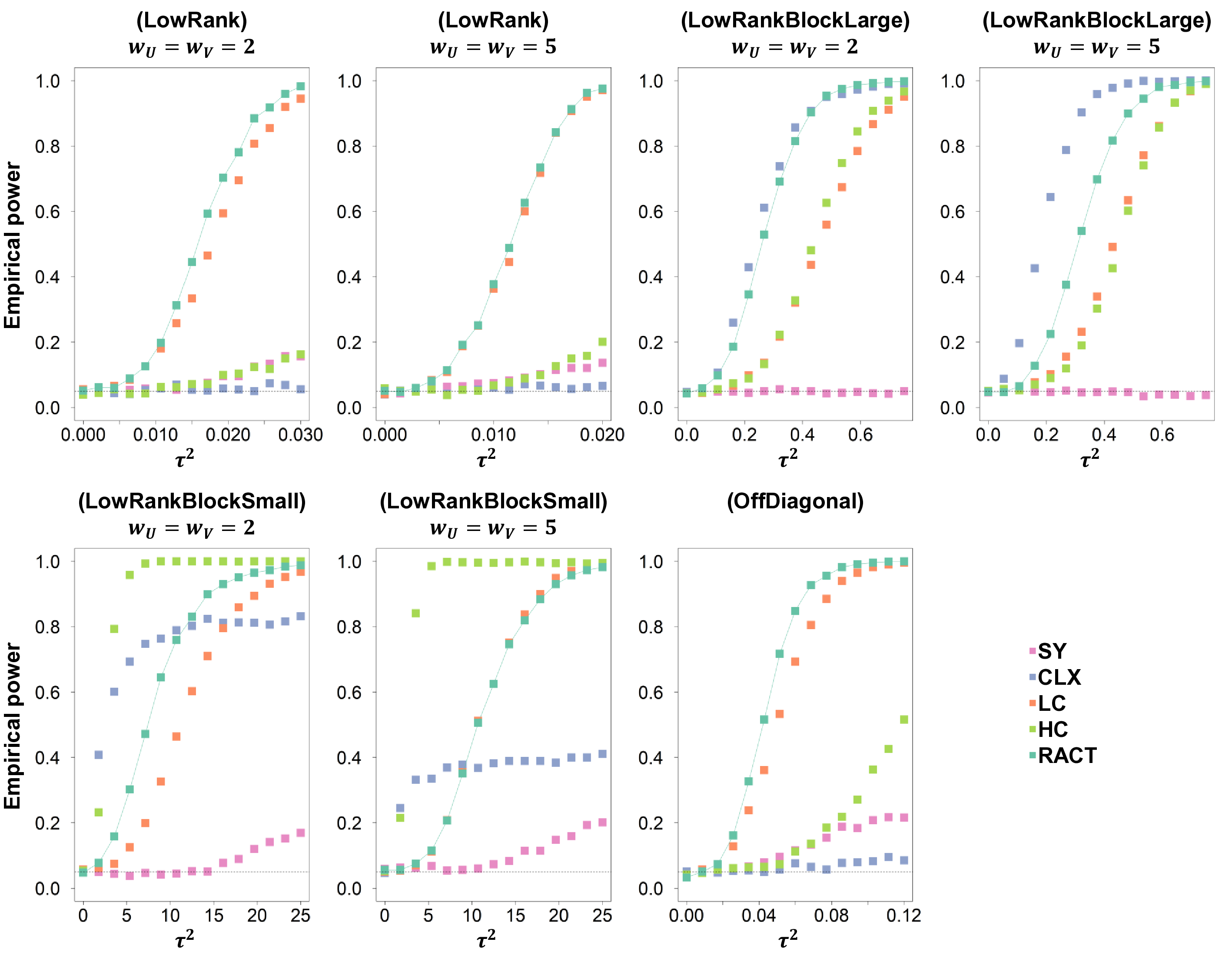}
	\caption{Empirical power curves for RACT and permutation-based versions of competing methods using simulated data. $n=50, p=250, \alpha=0.05$ for all simulations. RACT performs well across all scenarios, whereas the performance of other methods appears more variable.} 
 \label{fig:simulation_compre}
\end{figure}

Figure \ref{fig:simulation_compre} shows the empirical power of competing methods, and the significant differences in relative performance across simulation settings. The results indicate that while RACT does not outperform other methods uniformly in all settings, in each setting it is close to, if not the most powerful method. This contrasts with methods such as CLX and LC whose test statistics perform very strongly in certain scenarios, but lack power in others. Relatedly, these results suggest that even if one knows in advance the form of $\Sigma_1-\Sigma_2$, it is difficult to ascertain which two-sample covariance testing method will maximize power. We see this in S2 where CLX, which is expected to be particularly powerful when a small number of entries of $\Sigma_1-\Sigma_2$ are non-zero,  maximizes power across all methods when $\Sigma_1-\Sigma_2$ has many non-zero entries. This points to the importance of RACT's adaptivity in a real data setting, where even if there is a prior hypothesis as to the general structure of $\Sigma_1-\Sigma_2$, it could still be hard to pinpoint the specific test statistic that achieves the highest test power. Additional simulations were run for S1 for increasing values of $w_U$ and $w_V$. We see RACT outperforms LC for $w_U=w_V=2,5$, but as the rank of the difference increases, LC outperforms for $w_U=w_V=10,25,50,100$. This aligns with what would be expected as LC's test statistic involves an estimate of the squared Frobenius norm of $\Sigma_1-\Sigma_2$ and hence the signal would increase as the number of non-zero singular values increases, whereas for RACT, the additional signal may not be included depending on the value of $K$. The relative performance of other methods as $w_U$ and $w_V$ increases is largely unchanged.

\section{Real data analysis}\label{sec:realdataanalysis}

We examine the performance of RACT, implemented by the min$p$ approach in Remark \ref{rem:tractminp}, on two separate datasets and compare it to the performance of the same permutation-based versions of the competing methods described in Section \ref{sec:sims_compare}. For both applications, we compare power when both samples are from different groups, for different sizes of subsamples from the full sample. When comparing RACT to other methods, as well as to individual Ky-Fan($k$) norms, we select $K$ as described in Section \ref{sec:kyfan} (we also provide a sensitivity analysis for different percentage cutoffs for selecting $K$ in Web Appendix C). 

\begin{figure}[t]
	\begin{center}
		\includegraphics[width=\textwidth]{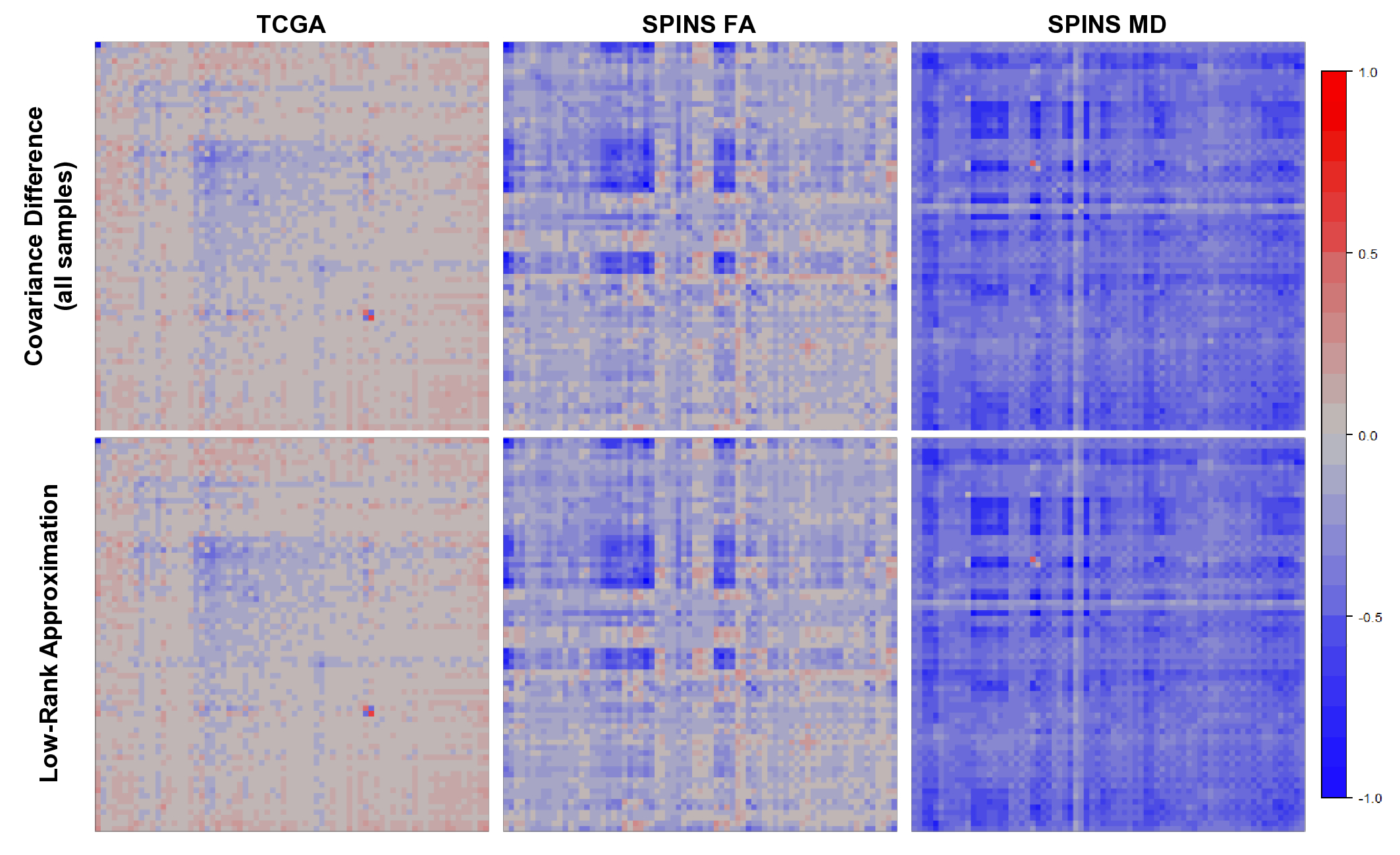}
	\end{center}
	\caption{Difference in covariance matrices between groups using all samples, and their low-rank approximations. For comparability across datasets, for each difference in covariance, entries were divided by the largest absolute value of all entries, so that all values fall in [-1,1]. Low-rank approximation is calculated via truncated SVD where the rank used is the smallest $k$ which maximized power for subsample size 20 (TCGA: $k=24$, SPINS FA: $k=10$, SPINS MD: $k=11$). TCGA features were ordered via hierarchical clustering for better visualization.}\label{fig:cov_mat_diff}
\end{figure}

\subsection{TCGA lung cancer data}

We analyze gene expression networks in two types of non-small cell lung cancer: (i) lung squamous cell carcinomas (LUSC) which is a common type of lung cancer, and (ii) lung adenocarcinomas (LUAD) which is a leading cause of cancer death. We access the gene expression levels for these tumors  using the \texttt{BiocManager} \texttt{R} package \citep{biocmanagerpackage}, for which $n_1=553$ LUSC tumor samples and $n_2=600$ LUAD tumor samples were available from 19962 protein-coding genes. We restrict our analysis to $p=72$ genes found in the KEGG pathway (Kyoto Encyclopedia of Genes and Genomes, \url{https://www.genome.jp/kegg/}) for non-small cell lung cancer which have non-zero variability among the samples we include, and transform the data by taking $\log_2(1+\text{count})$. Finally, within each tumor type, we regress out age and sex from the data, and then use the residuals for our evaluations.

\begin{figure}[h]
	\begin{center}
		\includegraphics[width=\textwidth]{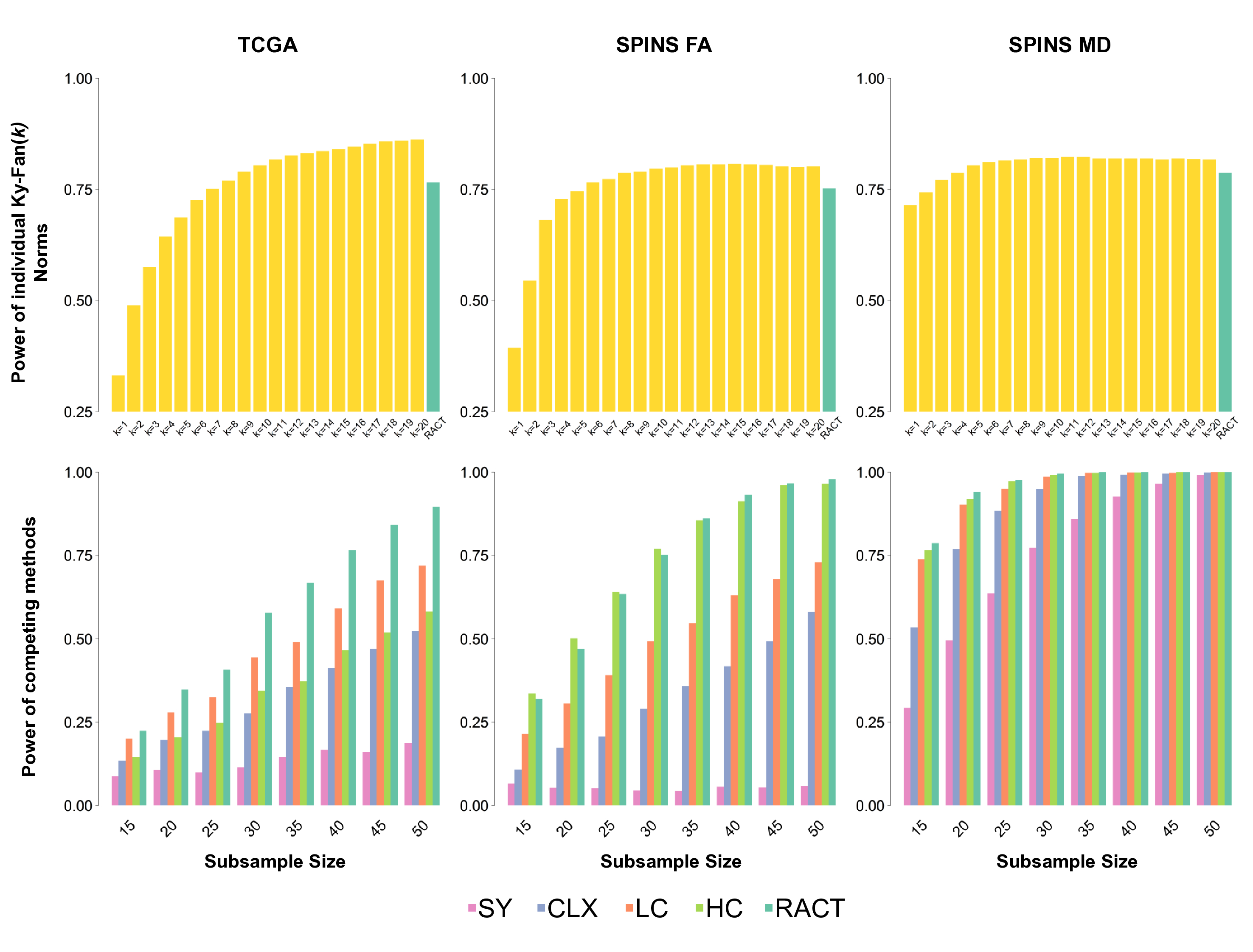}
	\end{center}
	\caption{First row: empirical power of individual Ky-Fan($k$) norms relative to RACT's power when using $K=50$. Subsample sizes presented: TCGA 40, SPINS FA 30, SPINS MD 15 (chosen so that power of RACT is approximately 75\%). Second row: empirical power for competing methods for various subsample sizes. }
	\label{fig:real_data_double}
\end{figure}

Using all samples, we see in the difference of covariance matrices between tumor types some evidence of a low-rank structure. The first singular value and the first 16 singular values, represent 21\% and 81\% of the total sum of all singular values respectively. This suggests that when testing the equality of these covariance matrices much of the signal can be found in a limited number of low-rank structures. In Figure \ref{fig:real_data_double} we examine the empirical power of various Ky-Fan($k$) norms for a fixed subsample size of $n_1=n_2={40}$. A rapid increase in power when $\alpha=0.05$ is seen as $k$ increases, before leveling out near $k=25$.  Although RACT is less powerful than the most powerful Ky-Fan($k$) norm, we do see that across subsample sizes RACT exhibits increased power relative to SY, CLX, LC, and HC. We observe that of the ten genes which load most heavily on the first singular vector of the difference in covariance, six appear on the OncoKB\textsuperscript{TM} cancer gene list \citep{suehnholz2024quantifying}: BAX, CDKN2A, ERBB2, HRAS, CDK4, and MAP2K2. CDKN2A is categorized as a tumor suppressor, whereas ERBB2, HRAS, CDK4, MAP2K2 are categorized as oncogenes.

\subsection{SPINS diffusion tensor imaging data}

The second dataset is from the Social Processes Initiative in Neurobiology of the Schizophrenia(s) (SPINS) study \citep{ds003011:1.2.0}. This dataset consists of diffusion tensor imaging (DTI) measurements of fractional anisotropy (FA) and mean diffusivity (MD). In the SPINS study most sites began with General Electric 3T (GE) scanners, and all ended with Siemens Prisma 3T (SP) scanners. In the below analysis we provide further evidence of inter-scanner covariance heterogeneity in the SPINS study.  Along the lines of \cite{10.1162/imag_a_00011} we use linear regression to remove the effect of age, age$^2$, gender, diagnosis, age $\times$ gender, and age $\times$ diagnosis, and then use these residuals to test for differences in the covariances. $p=73$ for both FA and MD, and in total we have $n_1=130$ from GE and $n_2=195$ from SP.

An examination of the difference in covariance matrices using all samples reveals a low-rank structure for both FA and MD measurements. For FA the first singular value and the sum of the first 19 singular values represent 30\% and 80\% of the total sum of all singular values. The low-rank structure is more pronounced for MD where the first singular value represents 59\% of the total sum, and the sum of the top 6 singular values represents 81\% of the total sum. Using a fixed subsample size of 30 and 15 for FA and MD respectively we see in Figure \ref{fig:real_data_double} for individual Ky-Fan($k$) norms power is maximized at $k=13$ for FA and $k=11$ for MD. Reflecting the lower-rank structure of the MD data, we see a less steep increase in power as $k$ increases relative to FA. Similar to the TCGA data we see relative to other methods RACT performs strongly. For SPINS-FA and SPINS-MD the low-rank structures associated with the first singular values for the differences in covariance we see Figure \ref{fig:cov_mat_diff}, appears similar to the scanner-specific differences which exist after applying harmonization of the competing methods presented in Figure 5 of \cite{10.1162/imag_a_00011}.

\subsection{Relative performance of competing methods}

Across TCGA, SPINS FA, and SPINS MD we note RACT's best relative performance appears for the TCGA dataset, where its empirical power is strictly higher than other methods across all subsample sizes tested. Also it is notable that RACT's performance is strong across all datasets whereas the performance of SY, CLX, and LC is more variable; this would be expected given RACT's adaptive test statistic as compared to the test statistics of other methods which will be more sensitive to the specific form the difference in covariance takes. We see in SPINS FA and SPINS MD that the method performing most similarly to RACT is HC, and we compare this relative performance more in Web Appendix C. 

\section{Discussion}\label{sec:discussion}

In this paper we propose a novel two-sample covariance testing method, which is able to improve power via leveraging low-rank structures commonly found in genomics and neuroimaging data. Underlying RACT is the use of the Ky-Fan($k$) matrix norm, which is novel in the setting of two-sample covariance testing. In Section \ref{sec:theory} we investigate the asymptotic properties of the Ky-Fan($k$) norm, and discover a delicate signal-to-noise trade-off which emerges for different values of $k$. This trade-off is reflected in our power simulations where the Ky-Fan($k$) norm which maximizes power is seen to differ between scenarios.

RACT utilizes an adaptive test statistic, composed of a series of individual Ky-Fan($k$) norms. RACT is able to adapt to the differences in low-rank structures, since for an appropriate $k$, the Ky-Fan($k$) norm captures most of the signal found in these low-rank differences.  However, we see in Figure \ref{fig:real_data_double} that in our real data applications, the Ky-Fan($k$) norms for very small $k$ have substantially reduced power (and in Figure \ref{fig:simulation} the operator norm generally does not maximize power). To solve this problem one could select a lower bound $K^*$ such that only Ky-Fan($k$) norms for $K^* \leq k\leq K$ are included, potentially choosing $K^*$ in a similar fashion to how $K$ is selected in Section \ref{sec:kyfan}.  Another potential extension would be to have $T_\text{RACT}$ take a maximum over test statistics other than Ky-Fan($k$) norms. If one possessed pre-existing knoweldge of the singular vectors of the difference of covariance one could use this to increase power. For example, if the signal was expected to be concentrated in a certain block, then one could include Ky-Fan($k$) norms calculated using only that block in $T_\text{RACT}$. On the other hand, if the differences in covariances were expected to have a banded structure then a test statistic similar to \cite{he2018high} could be included. Also, instead of considering a low-rank structure in covariance differences, considering element-wise sparse structure in covariance differences may yield higher power, which we believe can still be incorporated in the min$p$ approach \citep{he2021asymptotically}. However, as we see in Section \ref{sec:sims_compare}, competing two-sample covariance testing methods appear to have more variable performance, and including these test statistics may introduce a test statistic with very low power into $T_\text{RACT}$.

Several statistical tests such as higher criticism \citep{tukey1953problem}, Simes-type procedures \citep{simes1986improved}, and max-type tests, consider a family of related test statistics. $T_\text{RACT}$ most closely resembles a max-type test where the test statistics exhibit a high degree of dependence; this contrasts with Simes-type and higher criticism procedures which assume the individual tests are independent. These other methods would be expected to have higher power when the signal is distributed relatively evenly among individual tests.

Section \ref{sec:simulationstudy} showed how the rank of $\Sigma_1-\Sigma_2$ had a significant effect on the relative power of the individual Ky-Fan($k$) norms. Generally we saw that the Ky-Fan($k$) norms which maximize power were those where $k$ is close to the rank of $\Sigma_1-\Sigma_2$. In biomedical data analysis, techniques such as principal component analysis are often used for dimension reduction. With this in mind, the power maximizing $k$ of RACT could be used to help guide the selection of how many principal components to include for downstream analysis.

The procedure to select $K$, and RACT's test statistic, are functions of sample covariance matrices. However, in the high-dimensional setting the sample covariance may be a poor estimate for the population covariance. A potential extension could involve using high-dimensional regularization techniques \citep{fan2016overview} to select a more optimal $K$, or to better estimate $\widehat{\Sigma}_1-\widehat{\Sigma}_2$, and hence improve power. As well, given certain regularization techniques are well-suited for specific covariance structures (e.g., sparse or factor model-based), the regularization used for $\widehat{\Sigma}_1-\widehat{\Sigma}_2$ could be chosen via pre-existing knowledge of the data, or test statistics based on different regularization techniques could be included in the set of test statistics $T_\text{RACT}$ takes a maximum over.

\section*{Software}

An implementation of the RACT method in the form of an \texttt{R} package can be found at \url{https://github.com/daveveitch/RACT}.

\section*{Funding}

DV was partially supported by the  University of Toronto’s Data Sciences Institute. YH was partially supported by the Wisconsin Alumni Research Foundation. JYP was partially supported by the Natural Sciences and Engineering Research Council of Canada (RGPIN-2022-04831) the University of Toronto’s Data Sciences Institute, the McLaughlin Centre, and the Connaught Fund. The computing resources were enabled in part by support provided by University of Toronto and the Digital Research Alliance of Canada.

\section*{Conflict of interest}

None declared. 

\section*{Data availability}

The TCGA data used in Section 5 is publicly available from the \texttt{BiocManager} R package \citep{biocmanagerpackage}. The imaging data from the SPINS study are available upon reasonable request to Dr. Aristotle Voineskos (Centre for Addiction and Mental Health) with the execution of necessary data use agreements.

{\singlespacing
\bibliographystyle{apacite} 
\bibliography{biometrics}
}

\newpage 

\appendix


\section*{\centering Web Appendix}

\section{Minimum $p$-value method}

Below we outline the minimum $p$-value approach from Section 2.4, following \cite{pan2014powerful}. Instead of taking $T_k=||\widehat\Sigma_1-\widehat\Sigma_2||_{(k)}$, we take the $p$-value corresponding to $T_k$ computed from the empirical null distribution constructed by permutation. Specifically, we compute
\begin{align*}
    p_k=\frac{1}{B+1}\left[1+\sum_{b=1}^B I\left(T_k^{(b)} \geq T_k\right) \right]
\end{align*}
so that the min$p$-based RACT statistic becomes
\begin{align*}
    T_{\text{RACT-min}p} = \min_{k\in\mathcal{K}}  p_k.
\end{align*}
Although the null distribution of $T_{\text{RACT-min}p}$ is not provided in an explicit form, we use the $T_{k}^{(b)}$ statistics to construct an empirical null distribution. First, we construct a $p$-value for each $T_k^{(b)}$ as follows:
\begin{align*}
    p_k^{(b)} = \frac{1}{B}\left[1+\sum_{b_1 \neq b} I\left(T_k^{(b_1)} \geq T_k^{(b)}\right) \right],
\end{align*}
which is used to construct $T_{\text{RACT-min}p}^{(b)} = {\min_{k\in\mathcal{K}} p_k^{(b)}}$. Therefore, the $p$-value for the RACT-min$p$ is computed by
\begin{align*}
    p_\text{RACT-min$p$}=\frac{1}{B+1}\left[1+\sum_{b=1}^B I\left(T_{\text{RACT-min}p}^{(b)} \leq T_\text{RACT-min$p$}\right) \right].
\end{align*}


\section{Proofs of results in Section 3}

\subsection{Proof of Proposition \ref{thm:permutationconsit}}

The following proof follows from  general conclusion of permutation test; see, e.g., (17.8) in \cite{lehmann2021testing}. 
Let $G$ denote the group containing all possible permutations, and let $\mathbf{X}=(X_i^{(s)}: i=1,\ldots, n_s; s=1,2)$ denote the vector of all observations.
Under $H_0$, $X_i^{(s)} \sim \mathcal{N}(0,\Sigma)$  independently for $i=1,\ldots, n_s$ and $s=1,2$. 
Thus, for any permutation $g\in G$,  $ g\mathbf{X}$ and $\mathbf{X}$ follow the same distribution under $H_0$. 
Therefore, $T^{(1)}_{\operatorname{RACT}},\ldots, T^{(B)}_{\operatorname{RACT}}$, and $ T_{\operatorname{RACT}}$ are exchangeable and continuous variables. 
Then the rank of $T_{\operatorname{RACT}}$, i.e., $1+\sum_{b=1}^B I( T^{(b)}_{\operatorname{RACT}}\geqslant T_{\operatorname{RACT}} )$ is uniformly distributed on $1,\ldots, B+1$ 
\citep{kuchibhotla2020exchangeability}. 
Therefore, 
\begin{align*}
    \mathbb{P}(p_{\operatorname{RACT}}\leqslant \alpha) =   \mathbb{P}(\mathrm{rank}(T_{\operatorname{RACT}})\leqslant (B+1)\alpha  ) = \frac{\lfloor (B+1)\alpha \rfloor }{B+1}  \leqslant \alpha. 
\end{align*}


\subsection{Proof of Theorem 1} 
We first derive  \eqref{eq:powerform1_2} in Theorem 1. 
We  write the singular value decompositions  $\widehat{\Delta}=\widehat{\Sigma}_1-\widehat{\Sigma}_2=\sum_{j=1}^p\widehat{\lambda}_j\widehat{u}_j\widehat{v}_j^{\top}$ and $\Delta_0=\Sigma_1-\Sigma_2 = \sum_{j=1}^p \lambda_ju_jv_j^{\top}$, where $(\widehat{\lambda}_j$,$\lambda_j)$ denote the singular values, and $(\widehat{u}_j$,$u_j)$ and $(\widehat{v}_j$,$v_j)$ denote the left and right singular vectors, respectively. 
Write 
\begin{align}
\widehat{\lambda}_{j}=&~\widehat{\lambda}_{ j} u_j^{\top}u_j  
=  \widehat{\lambda}_{j} u_j^{\top}(\widehat{u}_j+u_j-\widehat{u}_j)\notag\\
=&~  u_j^{\top}\widehat{\Delta}\widehat{v}_j+\widehat{\lambda}_{ j} u_j^{\top}(u_j-\widehat{u}_j)\quad \  (  \text{by }  \widehat{\lambda}_{j} u_j^{\top} \widehat{u}_j = u_j^{\top} \widehat{\lambda}_j\widehat{u}_j = u_j^{\top}\widehat{\Delta}\widehat{v}_j)\notag\\
=&~u_{j}^{\top}\widehat{\Delta}(v_j+\widehat{v}_j -v_j) - \widehat{\lambda}_{ j} u_j^{\top}(\widehat{u}_j-u_j)\notag\\ 
=&~ u_{j}^{\top}\widehat{\Delta}v_j + u_j^{\top}\widehat{\Delta}(\widehat{v}_j-v_j)-u_j^{\top}\widehat{\lambda}_{j}(\widehat{u}_j -u_j ).  \label{eq:lambdahatdecomp} 
\end{align}
Since $T_k-\|{\Sigma}_1-{\Sigma}_2\|_{(k)} = \sum_{j=1}^k(\widehat{\lambda}_j-\lambda_j)$, 
by \eqref{eq:lambdahatdecomp}  and Slutsky's theorem, it suffices to show that 
\begin{align}
\frac{\sqrt{n}}{w_{1:k}}|u_j^{\top}\widehat{\Delta}(\widehat{v}_j-v_j)-u_j^{\top}\widehat{\lambda}_{j}(\widehat{u}_j -u_j ) | =   o_p(1), \label{eq:residualhat_op1}\\
 \frac{\sqrt{n}}{w_{1:k}}\sum_{j=1}^k    ( u_{j}^{\top}\widehat{\Delta}v_j  -\lambda_j)  \xrightarrow{d} \mathcal{N}(0,1), 
\label{eq:joint_limit_goal_1}
\end{align}
which are proved in Section \ref{sec:pftwosamlimitsv} below. 


Second, for \eqref{eq:consistent_power1} in Theorem 1,
 \eqref{eq:lambdahatdecomp}--\eqref{eq:joint_limit_goal_1} imply $T_k-\|\Sigma_1-\Sigma_2\|_{(k)} = o_P(1)$. 
Moreover, Equation \eqref{eq:ract_lower_power} holds  as  $  \mathbb{P}_{H_A}(T_{\mathrm{RACT}} \geqslant t)\geqslant  \mathbb{P}_{H_A}(T_{k} \geqslant t_k)$ for $k\in \mathcal{K}$ by the construction of $T_{\mathrm{RACT}}$.  
Therefore, by \eqref{eq:ract_lower_power}, 
\begin{align} 
  \mathbb{P}_{H_A}(T_{\mathrm{RACT}} \geqslant t) \geqslant   \mathbb{P}_{H_A}(T_k\geqslant t_k) = &~    \mathbb{P}_{H_A}(T_{k}-\|\Sigma_1-\Sigma_2\|_{(k)} \geqslant o(1)- \|\Sigma_1-\Sigma_2\|_{(k)}), \label{eq:pha_tk}
\end{align}
where the second equation holds by $t_k=o(1)$ for at least one of $k\in \mathcal{K}$; see   Lemma \ref{lm:tk_order}.
Since $\|\Sigma_1-\Sigma_2\|_{(k)}$ is bounded away from zero under Condition \ref{cond:regularity}, we know $\eqref{eq:pha_tk}\to 1$ as $n\to \infty$, giving   \eqref{eq:consistent_power1}.

\subsubsection{Proof of \eqref{eq:residualhat_op1} and \eqref{eq:joint_limit_goal_1} } \label{sec:pftwosamlimitsv}

As $\widehat{\Delta}$ and $\Delta_0$ are symmetric, we know $\widehat{u}_j=\pm \widehat{v}_j$ and $u_j=\pm v_j$. 
As the sign can be adjusted, we assume, without loss of generality that $v_j^{\top}\widehat{v}_j\geqslant 0$ below. 
Let $\widehat{\Lambda}_j$ and $\Lambda_j$ represent eigenvalues of $\widehat{\Delta}$ and $\Delta_0$, respectively,  and by matrix properties, they satisfy   $|\Lambda_j|=\lambda_j$, $|\widehat{\Lambda}_j|=\widehat{\lambda}_j$, 
\begin{align}\label{eq:singsign_prop}
    \widehat{v}_j=\widehat{u}_j\mathrm{sign}(\widehat{\Lambda}_j)\quad \text{ and }\quad v_j = u_j \mathrm{sign}(\Lambda_j). 
\end{align}

\paragraph{Proof of \eqref{eq:residualhat_op1}.}

As $U_k=[u_1,\ldots, u_k]$ equals $V_k=[v_1,\ldots, v_k]$ up to sign flips of columns, we have $U_kV_k^{\top}=V_kU_k^{\top}$, and then for $s=1,2$, 
\begin{align*}
 \mathrm{tr}\{(U_k^{\top}\Sigma_sV_k)^2\}   =\mathrm{tr}( U_k^{\top}\Sigma_s U_k V^{\top}_k\Sigma_s V_k)\geqslant \lambda_{\min}(U_k^{\top}\Sigma_s U_k)\, \lambda_{\min}(V_k^{\top}\Sigma_s V_k) \geqslant \lambda_{\min}^2(\Sigma_s).
\end{align*}
where $\lambda_{\min}(\cdot)$ represents the minimum singular value, and the inequalities are obtained as $\Sigma_s$ is positive definite, and $U_k$ and $V_k$ are orthonormal matrices. 
Therefore, $w_{1:k}^2=2\sum_{s=1}^2r_s\mathrm{tr}\{(U_k^{\top}\Sigma_sV_k)^2\}\geqslant 2\sum_{s=1}^2 \lambda_{\min}^2(\Sigma_s)$, and then  $1/w_{1:k}^2=O(1)$ under Condition \ref{cond:regularity}. 
To prove \eqref{eq:residualhat_op1}, it then suffices to show $ \sqrt{n}|u_j^{\top}\widehat{\Delta}(\widehat{v}_j-v_j)-u_j^{\top}\widehat{\lambda}_{j}(\widehat{u}_j -u_j ) |  =o_p(1)$. 

By $u_j^{\top}u_j=v_j^{\top}v_j=1$, 
\begin{align}
 \sqrt{n}|u_j^{\top}\widehat{\Delta}(\widehat{v}_j-v_j)-u_j^{\top}\widehat{\lambda}_{j}(\widehat{u}_j -u_j ) |  
=&~\sqrt{n}|u_j^{\top}\widehat{\Delta}(\widehat{v}_j-v_j)-\widehat{\lambda}_{j}(u_j^{\top}\widehat{u}_j -v_j^{\top}v_j ) | \leqslant I+II, \notag
\end{align}
where we define
\begin{align*}
    I = &~ \sqrt{n}|u_j^{\top}\widehat{\Delta}(\widehat{v}_j-v_j)-\widehat{\lambda}_{j}v_j^{\top}(\widehat{v}_j-v_j) |\quad \text{ and }\quad 
II=\sqrt{n}\widehat{\lambda}_j|v_j^{\top}\widehat{v}_j -u_j^{\top}\widehat{u}_j |.
\end{align*}
We next show $I=o_P(1)$ and $II=0$ almost surely. 
First, 
\begin{align*}
    I = &~ \sqrt{n}|(u_j^{\top}\widehat{\Delta}-\widehat{\lambda}_{j}v_j^{\top})(\widehat{v}_j-v_j) | \\
    =&~  \sqrt{n}|(u_j^{\top}\widehat{\Delta}- u_j^{\top}\Delta_0-\widehat{\lambda}_{j}v_j^{\top} + \lambda_j v_j)(\widehat{v}_j-v_j) | \quad \quad (\text{by } u_j^{\top}\Delta_0 =\lambda_j v_j^{\top}) \\
    =&~ \sqrt{n}  |\{ u_j^{\top}(\widehat{\Delta}-\Delta_0) - v_j^{\top} (\widehat{\lambda}_j-\lambda_j) \} (\widehat{v}_j-v_j) |  
    \leqslant  2\sqrt{n}\|\widehat{\Delta}-\Delta_0\|_{op}  \|\widehat{v}_j-v_j\|_2,
\end{align*}
where $\|\cdot\|_{op}$ represents matrix operator norm, and in the last inequality, we use Weyl's inequality, $|\widehat{\lambda}_j-\lambda_j|\leqslant  \|\widehat{\Delta}-\Delta_0\|_{op}$. 
By Davis-Kahan theorem \citep[e.g., ][Theorem 2]{yu2015useful}, we have 
\begin{align*}
 \|\widehat{v}_j-v_j\|\leqslant  \frac{2 \|\widehat{\Delta}-\Delta_0\|_{op}}{\min\{\Lambda_j-\Lambda_{j+1},\Lambda_{j-1} - \Lambda_j\}}.
\end{align*}
Therefore, 
\begin{align}\label{eq:deltahat_expand}
    I \leqslant \frac{4\sqrt{n} \|\widehat{\Delta}-\Delta_0\|_{op}^2}{\min\{\Lambda_j-\Lambda_{j+1},\ \Lambda_{j-1} - \Lambda_j\}}. 
\end{align}
By the definition of sample covariances, we have 
\begin{align*}
   \widehat{\Delta}= &~\widehat{\Sigma}_1-\widehat{\Sigma}_2 
   =\sum_{s\in \{1,2\}} (-1)^{s+1}\left(1+\frac{1}{n_s-1} \right)\left(\frac{1}{n_s} \sum_{i=1}^{n_s}X_i^{(s)}X_i^{(s)\top}- \overline{X}^{(s)} \overline{X}^{(s)\top} \right), 
\end{align*}
where we let $\overline{X}^{(s)} = \sum_{i=1}^{n_s}X_i^{(s)}/n_s$ denote the sample mean for the $s$-th sample for $s\in  \{1,2\}$. 
To facilitate the analysis, we define intermediate terms
\begin{align*}
    \widehat{\Delta}_{\mu}=&~\sum_{s\in \{1,2\}} (-1)^{s+1}\left(1+\frac{1}{n_s-1} \right)\frac{1}{n_s} \sum_{i=1}^{n_s}X_i^{(s)} X_i^{(s)\top},\\
    \Delta_{0,1}=&~ \sum_{s\in \{1,2\}} (-1)^{s+1}\left(1+\frac{1}{n_s-1} \right)\Sigma_s, \\
    \Delta_{0,2}=&~ \sum_{s\in \{1,2\}} (-1)^{s+1} \frac{(-1)}{n_s-1}  \Sigma_s.  
\end{align*}
We have $\Delta_0=\Sigma_1-\Sigma_2=\Delta_{0,1}+\Delta_{0,2}$. Then 
\begin{align}
\|\widehat{\Delta}-\Delta_0\|_{op}=&~ \|\widehat{\Delta}-\widehat{\Delta}_{\mu}+\widehat{\Delta}_{\mu}-\Delta_{0,1}-\Delta_{0,2}\| _{op} \notag \\
\leqslant &~ \|\widehat{\Delta}-\widehat{\Delta}_{\mu}\|_{op}+\|\widehat{\Delta}_{\mu}-\Delta_{0,1}\|_{op}+\|\Delta_{0,2}\|_{op}. \label{eq:deltahat_op_error}
\end{align}
By the definitions above, 
\begin{align}
     \|\widehat{\Delta}-\widehat{\Delta}_{\mu}\|_{op}=&~
     \left\| \sum_{s\in \{1,2\}} (-1)^{s} \left(1+\frac{1}{n_s-1} \right) \, \overline{X}^{(s)} \overline{X}^{(s)\top}  \right\|_{op}  
       \leqslant 2\sum_{s\in \{1,2\}} \|\overline{X}^{(s)}\|_{2}^2 = O_p(p/n), \notag \\[2pt]
    \|\Delta_{0,2}\|_{op} \leqslant &~ \frac{2}{n_1}\|\Sigma_1\|_{op}+ \frac{2}{n_2}\|\Sigma_2\|_{op},\\[2pt]
    \|\widehat{\Delta}_{\mu}-\Delta_{0,1}\|_{op} \leqslant &~ \sum_{s\in \{1,2\}} \left\|\frac{1}{n_s} \sum_{i=1}^{n_s} X_i^{(s)} X_i^{(s)\top} - \Sigma_s \right\|_{op}=O_p\left[\, (\|\Sigma_1\|_{op}+\|\Sigma_2\|_{op})\, \sqrt{\frac{p}{n}}\, \right], \label{eq:delta_cov_error}
\end{align}
where \eqref{eq:delta_cov_error} follows by existing conclusions on the covariance estimation, see, e.g., \cite{wainwright2019high}, Example 6.3.
Combining all the equations between \eqref{eq:deltahat_expand} and \eqref{eq:delta_cov_error}, and letting  $\delta_{1:k}=\min_{1\leqslant j\leqslant k} \{\Lambda_j-\Lambda_{j+1},\ \Lambda_{j-1} - \Lambda_j\}$ denote the minimum eigen gap, we have 
\begin{align*}
I= &~ \frac{k}{\delta_{1:k}} O_p\left\{\left(\|\Sigma_1\|_{op}^2+\|\Sigma_2\|_{op}^2+\frac{p}{n}\right)\frac{p}{\sqrt{n}} + \frac{\|\Sigma_1\|_{op}^2+\|\Sigma_2\|_{op}^2}{n^{3/2}} \right\}=o_p(1) 
\end{align*} 
when $\|\Sigma_1\|_{op}^2+\|\Sigma_2\|_{op}^2$ is finite and $p = o(\sqrt{n})$. 

Second, by matrix properties of singular vectors, we know $II=0$ if  $v_j^{\top}\widehat{v}_j =u_j^{\top}\widehat{u}_j$, which is equivalent to $\mathrm{sign}(\widehat{\Lambda}_j)=\mathrm{sign}(\Lambda_j)$ by \eqref{eq:singsign_prop}.
Moreover, we next argue that $\mathrm{sign}(\widehat{\Lambda}_j)=\mathrm{sign}(\Lambda_j)$ holds if 
\begin{align}\label{eq:absbound}
    |\widehat{\Lambda}_j-\Lambda_j|\leqslant |\Lambda_j|/2. 
\end{align}
In particular, if $\Lambda_j>0$, \eqref{eq:absbound} implies $\widehat{\Lambda}_j-\Lambda_j\in [-\Lambda_j/2,\Lambda_j/2]$, and then $\widehat{\Lambda}_j\geqslant \Lambda_j/2>0$ has the same sign as $\Lambda_j>0$; if $\Lambda_j<0$, \eqref{eq:absbound} implies $\widehat{\Lambda}_j-\Lambda_j\in [\Lambda_j/2, -\Lambda_j/2]$ and then $\widehat{\Lambda}_j\leqslant \Lambda_j/2<0$ has the same sign as $\Lambda_j<0$.  
Therefore, 
\begin{align*}
  \mathbb{P}( II =0 )=\mathbb{P}\left\{\mathrm{sign}(\widehat{\Lambda}_j)=\mathrm{sign}(\Lambda_j)\right\}& \geqslant  
  \mathbb{P}(|\widehat{\Lambda}_j-\Lambda_j|\leqslant |\Lambda_j|/2)  \geqslant \mathbb{P}(\|\widehat{\Delta}-\Delta_0\|_{op}\leqslant |\Lambda_j|/2 )\to 1,
\end{align*}
where the second inequality follows by 
by Weyl's inequality, and the last convergence follows by $\|\widehat{\Delta}-\Delta_0\|_{op}=o_p(1)$ given \eqref{eq:deltahat_op_error}--\eqref{eq:delta_cov_error}.

\paragraph{Proof of \eqref{eq:joint_limit_goal_1}.} 
By $\lambda_j=u_j^{\top}\Delta_0 v_j$  and $\Delta_0=\Delta_{0,1}+\Delta_{0,2}$, 
we decompose 
\begin{align*}
  u_j^{\top}\widehat{\Delta}v_j-\lambda_j   =&~u_j^{\top}(\widehat{\Delta}_{\mu}-\Delta_{0,1})v_j-u_j^{\top}\Delta_{0,2}v_j+u_j^{\top}(\widehat{\Delta}-\widehat{\Delta}_{\mu})v_j\\
  =&~ u_j^{\top}(\widehat{\Delta}_{\mu}-\Delta_{0,1})v_j+ O_p(p/n), 
\end{align*}
where the second equation follows by 
\eqref{eq:deltahat_op_error}--\eqref{eq:delta_cov_error} such that
\begin{align*}
     |u_j^{\top}\Delta_{0,2}v_j|\leqslant \|\Delta_{0,2}\|_{op} \lesssim \frac{\|\Sigma_{1}\|_{op}+\|\Sigma_{2}\|_{op}}{n},\quad \text{and} \quad 
        |u_j^{\top}(\widehat{\Delta}-\widehat{\Delta}_{\mu})v_j|\leqslant  \|\widehat{\Delta}-\widehat{\Delta}_{\mu}\|_{op}=O_p(p/n). 
\end{align*}
As $\EXPT\{X_i^{(s)} X_i^{(s)\top} - \Sigma_s \} =0$, 
\begin{align}\label{eq:leadingdeltanorm}
\sqrt{n}\sum_{j=1}^k u_{j}^{\top}(\widehat{\Delta}_{\mu}-\Delta_{0,1})v_j=  \sqrt{n}\sum_{s=1}^2  (-1)^{s+1}\frac{1}{n_s-1} \sum_{i=1}^{n_s} \sum_{j=1}^k u_j^{\top} \big( X_i^{(s)} X_i^{(s)\top} - \Sigma_s \big)v_j
\end{align}  
is a summation of independent and mean-zero terms.
By the central limit theorem and Slutsky's theorem, to finish the proof of  \eqref{eq:joint_limit_goal_1}, it remains to show $ \mathrm{Var}\{\eqref{eq:leadingdeltanorm}\}/w_{1:k}^2\to 1$.

In particular, 
\begin{align}\label{eq:varsum}
    \mathrm{Var}\{\eqref{eq:leadingdeltanorm}\}=\sum_{s=1}^2r_s\{1+o(1)\}\sum_{j,l=1}^k\mathrm{cov}\left\{ u_j^{\top}X_i^{(s)}X_i^{(s)\top}v_j,\, u_l^{\top}X_i^{(s)}X_i^{(s)\top}v_l\right\},
\end{align} 
where $r_s=n/n_s$ is defined in Condition \ref{cond:regularity}. 
Given $s\in \{1,2\}$, for the simplicity of notation, let 
$\xi=(\xi_1,\ldots, \xi_p)^{\top}=\Sigma_{s}^{-1/2} X_i^{(s)}$ and $a_j=(a_{j1},\ldots, a_{jp})^{\top} = \Sigma_{s}^{1/2}u_{j}$. 
Then $\xi \sim \mathcal{N}(0, I_p)$, and by $u_{j} = \mathrm{sign}(\Lambda_j) v_j$, 
\begin{align*}
&~\mathrm{cov}\left\{u_j^\top  X_i^{(s)}  X_i^{(s)\top}  v_j,\  u_l^\top   X_i^{(s)} X_i^{(s)\top} v_l\right\} \times   \mathrm{sign}(\Lambda_j) \mathrm{sign}(\Lambda_l) \notag\\ 
= &~\mathrm{cov}\left\{v_j^\top  X_i^{(s)}  X_i^{(s)\top} u_j,\  u_l^\top   X_i^{(s)} X_i^{(s)\top} u_l\right\}  \notag\\ 
=&~\EXPT \left\{ (a_j^{\top}\xi)^2 \times (a_l^{\top}\xi)^2 \right\} -\EXPT\left\{ (a_j^{\top}\xi)^2 \right\}\times \EXPT\left\{ (a_l^{\top}\xi)^2 \right\}  \\
=&~ \sum_{t_1,t_2,t_3,t_4=1}^p \EXPT\left(\prod_{m=1}^2a_{jt_m}\xi_{t_m} \prod_{m=3}^4a_{lt_m}\xi_{t_m} \right) - a_j^{\top}a_j \times   a_l^{\top}a_l \\
=&~ \sum_{t=1}^p a_{jt}^2a_{lt}^2\EXPT(\xi_t^4) + \sum_{1\leq t_1\neq t_3\leq p}a_{jt_1}^2a_{lt_3}^2\EXPT (\xi_{t_1}^2\xi_{t_3}^2) +2 \sum_{1\leq t_1\neq t_2\leq p}a_{jt_1}a_{lt_1} a_{jt_2}a_{lt_2} \EXPT(\xi_{t_1}^2)\EXPT(\xi_{t_2}^2)\\
=&~ \left(\sum_{t=1}^pa_{jt}^2\right)  \left(\sum_{t=1}^pa_{lt}^2\right) + 2 \left(\sum_{t=1}^p a_{jt}a_{lt}\right)^2 - a_j^{\top}a_j\times a_l^{\top}a_l \\
=&~2 (a_j^{\top}a_l)^2 = 2 (u_j^{\top}\Sigma_s u_l)^2. 
\end{align*}
Therefore, 
\begin{align*}
\eqref{eq:varsum}= &~2\sum_{s=1}^2 r_s\{1+o(1)\} \sum_{j,l=1}^k  (u_j^{\top}\Sigma_s u_l)^2 \, \mathrm{sign}(\Lambda_j) \mathrm{sign}(\Lambda_l) \\
    =&~2 \sum_{s=1}^2 r_s\{1+o(1)\}\sum_{j,l=1}^k(u_j^{\top}\Sigma_s v_l) (v_j^{\top}\Sigma_s u_l) \quad (\text{by } v_j=u_j\mathrm{sign}(\Lambda_j) \text{ in } \eqref{eq:singsign_prop})\\
        =&~2 \sum_{s=1}^2 r_s\{1+o(1)\}\mathrm{tr}(U_k^{\top}\Sigma_sV_kU_k^{\top}\Sigma_sV_k), 
\end{align*}
 satisfying  $\eqref{eq:varsum}/w_{1:k}^2\to 1 $.

\subsection{Proof of Proposition 1} 
By the definition of $\mathrm{SNR}_k$, 
$\mathrm{SNR}_{k_2}\geq \mathrm{SNR}_{k_1}$ if and only if 
\begin{align*}
&~~ \|\Sigma_1-\Sigma_2\|_{(k_2)} {\omega_{1:k_1}} \geq \|\Sigma_1-\Sigma_2\|_{(k_1)} {\omega_{1:k_2}}  \notag\\
 \Leftrightarrow &~~ \frac{ \|\Sigma_1-\Sigma_2\|_{(k_2)} -\|\Sigma_1-\Sigma_2\|_{(k_1)}  }{\|\Sigma_1-\Sigma_2\|_{(k_1)} } \geq \frac{{\omega_{1:k_2}}}{{\omega_{1:k_1}}}-1 \notag\\
 \Leftrightarrow &~~ \beta_{k_1,k_2}\geq \sqrt{\gamma_{k_1,k_2}+1}-1. 
\end{align*}

\subsection{Threshold order $H_0$ } \label{sec:threshold_order}

\begin{lemma} \label{lm:tk_order}
Assume the data distribution satisfies Condition \ref{cond:regularity},
 and finite values of $k$ is used in $\mathcal{K}$. 
If there exists a fixed constant $c_0$ such  $\mathbb{P}_{H_0}(T_{\mathrm{RACT}} \geqslant t) \in [c_0\alpha, \alpha]$ given a significance level $\alpha\in (0,1)$, 
  there exists a constant $C>0$ 
  and at least one of $k\in \mathcal{K}$ 
  such that $0\leqslant  t_k \leqslant C \sqrt{p/n}=o(1)$. 
\end{lemma}

\paragraph{Proof}

As $T_k\geqslant 0$, if $t_k<0$ for one of $k\in \mathcal{K}$, we know $ T_k\geqslant t_k$ always holds and then $\mathbb{P}_{H_0}(T_{\mathrm{RACT}} > t)\geqslant \max_{k\in \mathcal{K}}\mathbb{P}_{H_0}(T_k\geqslant t_k) = 1$, which conflicts with $\mathbb{P}_{H_0}(T_{\mathrm{RACT}} > t)\leqslant \alpha <1$ in the assumption. Therefore, we must have $t_k\geqslant 0$ for all $k\in \mathcal{K}$.  

We next prove that $t_k\leqslant C\sqrt{p/n}$ for at least one of $k\in \mathcal{K}$. 
By our definition of $T_{\mathrm{RACT}}$, we know $\mathbb{P}_{H_0}(T_{\mathrm{RACT}} > t)  = \mathbb{P}_{H_0}(\cup_{k \in \mathcal{K}}\{T_k>t_k\} ) \leqslant \sum_{k \in \mathcal{K}} \mathbb{P}_{H_0}(T_{k} > t_k)$. 
Therefore,  under the assumption of Lemma \ref{lm:tk_order},
\begin{align}
    \sum_{k \in \mathcal{K}} \mathbb{P}_{H_0}(T_{k} > t_k) \geqslant c_0\alpha. \label{eq:sumlower_alpha}
\end{align}
If $t_k> C\sqrt{p/n}$ for all $k\in \mathcal{K}$, we know
\begin{align} \label{eq:sumlower_alpha2}
\sum_{k \in \mathcal{K}} \mathbb{P}_{H_0}(T_k> t_k) \leqslant  &~  \sum_{k \in \mathcal{K}} \mathbb{P}_{H_0}(T_k> C\sqrt{p/n}). 
\end{align} 
When $\Sigma_1=\Sigma_2$, 
since $ T_k=\|\widehat{\Sigma}_1-\widehat{\Sigma}_2\|_{(k)} =\|(\widehat{\Sigma}_1-\Sigma_1 )-(\widehat{\Sigma}_2-\Sigma_2)\|_{(k)} \leqslant k \sum_{s=1}^2\|\widehat{\Sigma}_s-\Sigma_s\|_{(1)}, $ we know 
\begin{align} \label{eq:tk_epsilon}
    \mathbb{P}_{H_0}\left(T_k>C\sqrt{\frac{p}{n}} \right)\leqslant &~  \mathbb{P}_{H_0}\left\{k \sum_{s=1}^2\|\widehat{\Sigma}_s-\Sigma_s\|_{(1)} > C \sqrt{\frac{p}{n}} \right\} \notag \\
    \leqslant &~\sum_{s=1}^2 \mathbb{P}_{H_0}\left\{ \|\widehat{\Sigma}_s-\Sigma_s\|_{(1)} > \frac{C }{2k} \sqrt{\frac{p}{n}}
    \right\}.
\end{align}
By Theorem 6.5 in \cite{wainwright2019high},  $ \|\widehat{\Sigma}_s-\Sigma_s\|_{(1)}/\|\Sigma_s\|_{(1)} = O_P( \sqrt{p/n})$. 
Therefore, there exists $C>0$ such that $\eqref{eq:tk_epsilon}< c_0\alpha/|\mathcal{K}|$ for all $k\in \mathcal{K}$, where $|\mathcal{K}|$ represents the size of $\mathcal{K}$ which is finite. Plugging this bound of \eqref{eq:tk_epsilon} into \eqref{eq:sumlower_alpha2} gives 
\begin{align*}
     \sum_{k \in \mathcal{K}}\mathbb{P}_{H_0}(T_k> C\sqrt{p/n})<c_0\alpha,
\end{align*}
which conflicts with \eqref{eq:sumlower_alpha}. 
Therefore, we must have $t_k\leqslant C\sqrt{p/n}$ for at least one of $k\in \mathcal{K}$.  

\begin{remark}
Lemma \ref{lm:tk_order} requires weak assumptions on the covariance structure $\Sigma_1=\Sigma_2$ under $H_0$ to establish an upper bound on  $t_k$ for one of $k\in \mathcal{K}$. As Section \ref{sec:empirical_ky_fan_k} demonstrates,  the distribution of $T_{\mathrm{RACT}}$ varies with respect to $\Sigma_1=\Sigma_2$. Therefore, a precise characterization of   $t$ and all $t_k$'s under general $H_0$ is challenging. Nevertheless, by $1 \leqslant T_k/T_1\leqslant  k$ and our numerical observations, we expect that the expectations and variances of $T_k$ for different finite $k$ values are likely to be of the same order. In that case, $t_k=t\sqrt{\mathrm{Var}_{H_0}[T_k]}+\mathrm{E}_{H_0}[T_k]$ for different $k$ are similar, and then we expect $t_k=o(1)$ for all $k\in \mathcal{K}$ too.  
\end{remark}




\section{Additional empirical results}

\subsection{Empirical distribution of $T_k$ under $H_0$}\label{sec:empirical_ky_fan_k}

In Web Figures \ref{fig:tract_distribution_01}, \ref{fig:tract_distribution_05}, \ref{fig:tract_distribution_10}, and \ref{fig:tract_distribution_50} we present the empirical distribution of our $T_k$ statistic for $k=1,5,10,50$ under $H_0$ that $\Sigma_1=\Sigma_2$ for various covariance structures. These figures demonstrate the way in which under $H_0$ the properties of the distribution of the $T_k$ can vary depending on what the population covariance is. We present results in the $n=1000,p=250$ setting for 1000 simulated datasets, for the following population covariance structures

\begin{enumerate}
    \item \textbf{(IID)}: $\Sigma_1=\Sigma_2=I$.
    \item \textbf{(LowRank) $w_U=w_V=2$}: Constructed in the way same as in Section \ref{sec:simulationsetup} of the main article with $\tau^2=0.5$ and the rank of the low-rank component equal to 2. One covariance matrix was randomly generated, and used for all 1000 datasets.
    \item \textbf{(LowRank) $w_U=w_V=5$}: Constructed in the same way as in Section \ref{sec:simulationsetup} of the main article with $\tau^2=0.5$ and the rank of the low-rank component equal to 5. One covariance matrix was randomly generated, and used for all 1000 datasets. 
    \item \textbf{(OffDiagonal)}: Constructed in the same way as in Section \ref{sec:simulationsetup} of the main article with $\tau^2=0.5$.
    \item \textbf{(AR)}: The $[r,s]$ entry of the population covariance $\Sigma$ is $\Sigma[r,s]=0.8^{|r-s|}$.
\end{enumerate}

To facilitate better visualization across different covariance structures, we standardize the empirical $T_k$ statistics. Letting $T_k^{(1)},\dots,T_k^{(1000)}$ represent the $T_k$ statistics across all 1000 datasets for some covariance structure, we calculate $T_k^{*(j)}=(T_k^{(j)}-\widehat{\mu}_k)/\widehat{\sigma}_k$ where $\widehat{\mu}_k,\widehat{\sigma}_k$ are the empirical mean and standard deviation of $T_k^{(1)},\dots,T_k^{(1000)}$, and then present the $T_k^{*(j)}$ in the figures below. Along the diagonals we present the empirical distribution of $T_k^{*(j)}$ for each covariance structure, and in the off-diagonals present QQ-plots comparing the $T_k^{*(j)}$ from two covariance structures. For example, the bottom left plot in Web Figure \ref{fig:tract_distribution_01} is a QQ-plot comparing the empirical distribution of the standardized $T_1$ statistic under the null for (IID) and (AR) covariance structures, with (IID) on the x-axis and (AR) on the y-axis. We can see from both the histograms and QQ-plots that the population covariance appears to influence the skewness and tail behaviour of the $T_k$ distributions.

\begin{figure}[H]
\centering
    \includegraphics[width=0.95\textwidth]{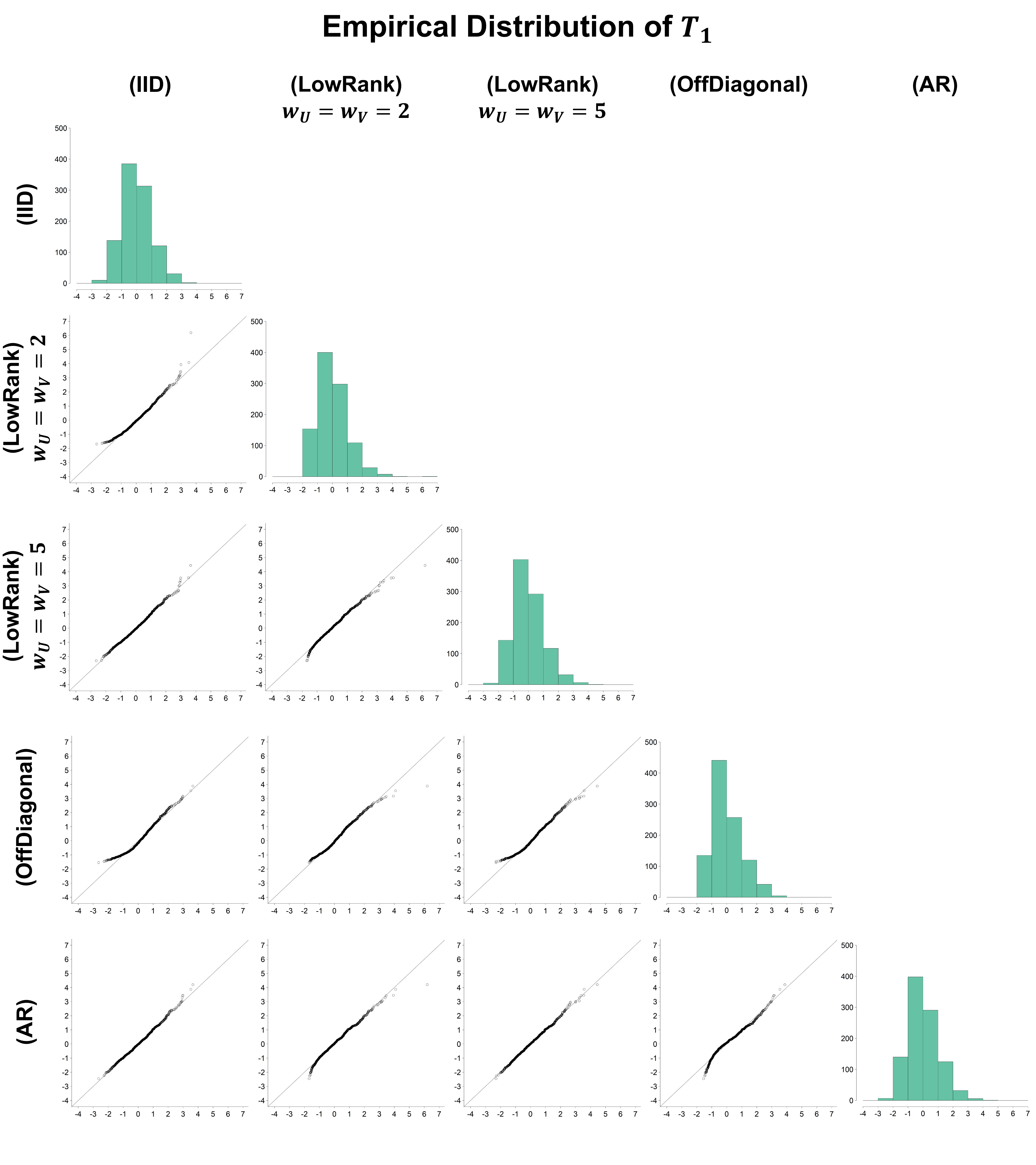}
     \caption{Empirical distribution of standardized $T_1$ under $H_0$ for various covariance structures in the $n=1000,p=250$ setting.}
     \label{fig:tract_distribution_01}
\end{figure}

\begin{figure}[H]
\centering
    \includegraphics[width=0.95\textwidth]{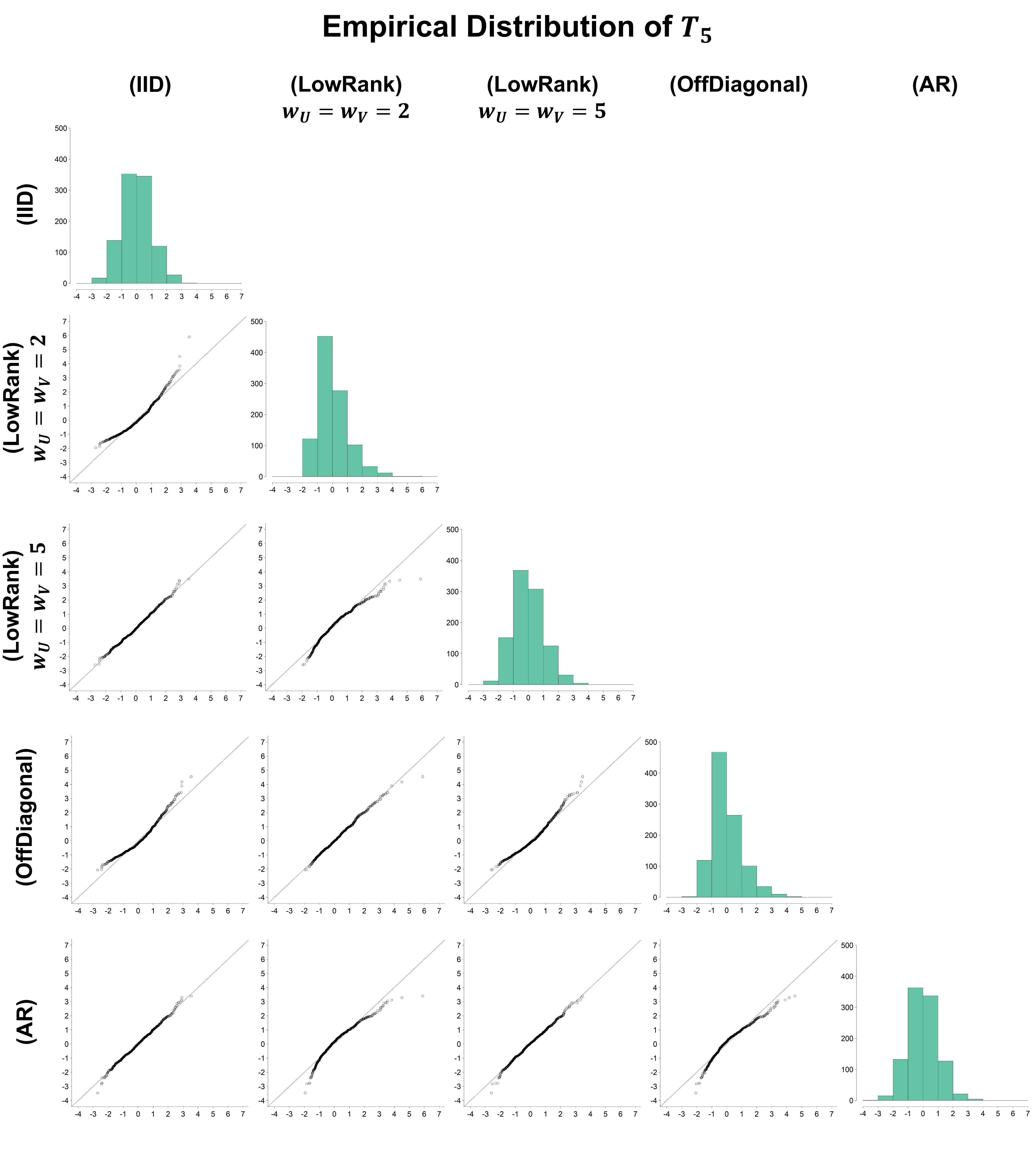}
     \caption{Empirical distribution of standardized $T_5$ under $H_0$ for various covariance structures in the $n=1000,p=250$ setting.}
     \label{fig:tract_distribution_05}
\end{figure}

\begin{figure}[H]
\centering
    \includegraphics[width=0.95\textwidth]{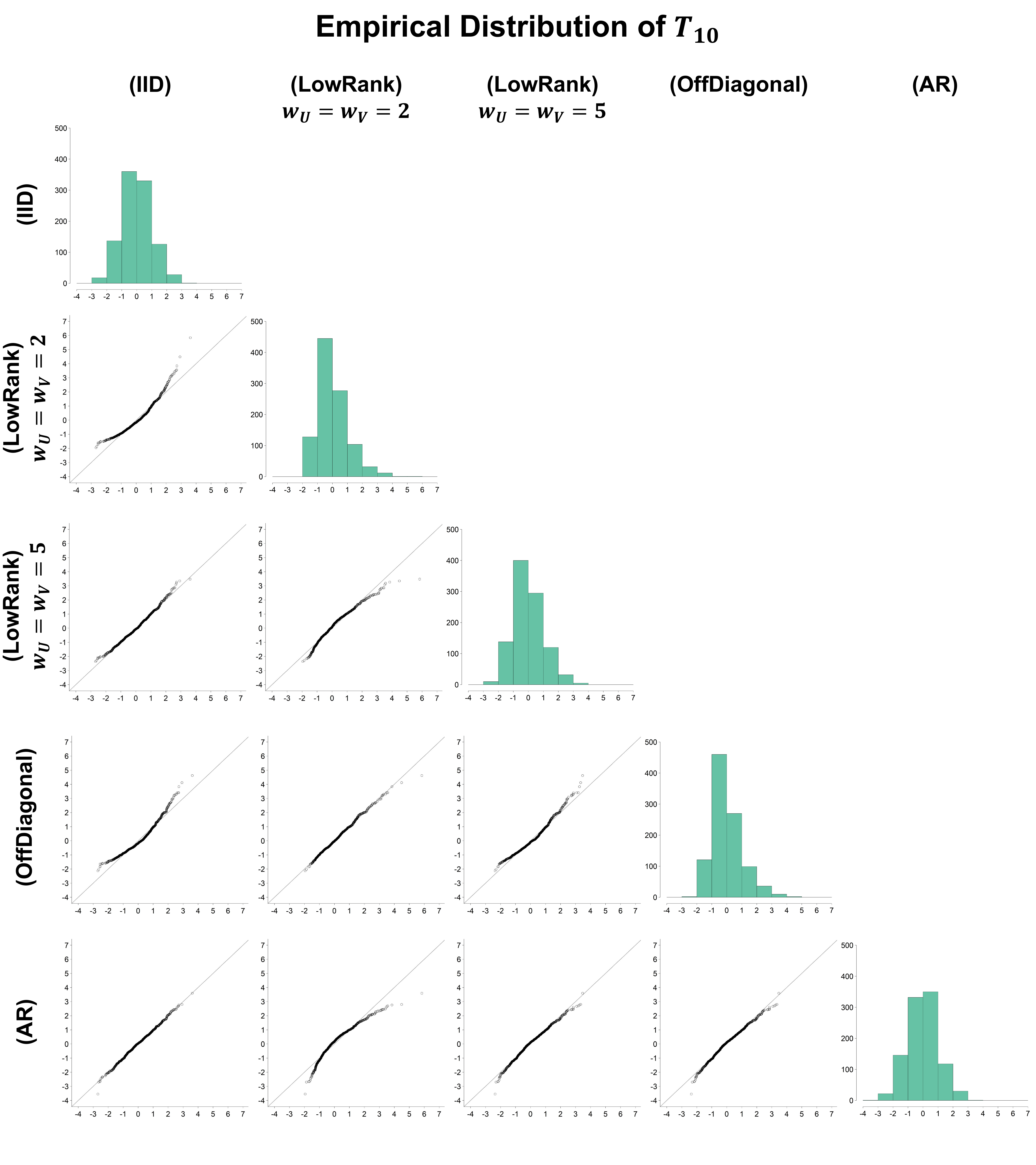}
     \caption{Empirical distribution of standardized $T_{10}$ under $H_0$ for various covariance structures in the $n=1000,p=250$ setting.}
     \label{fig:tract_distribution_10}
\end{figure}

\begin{figure}[H]
\centering
    \includegraphics[width=0.95\textwidth]{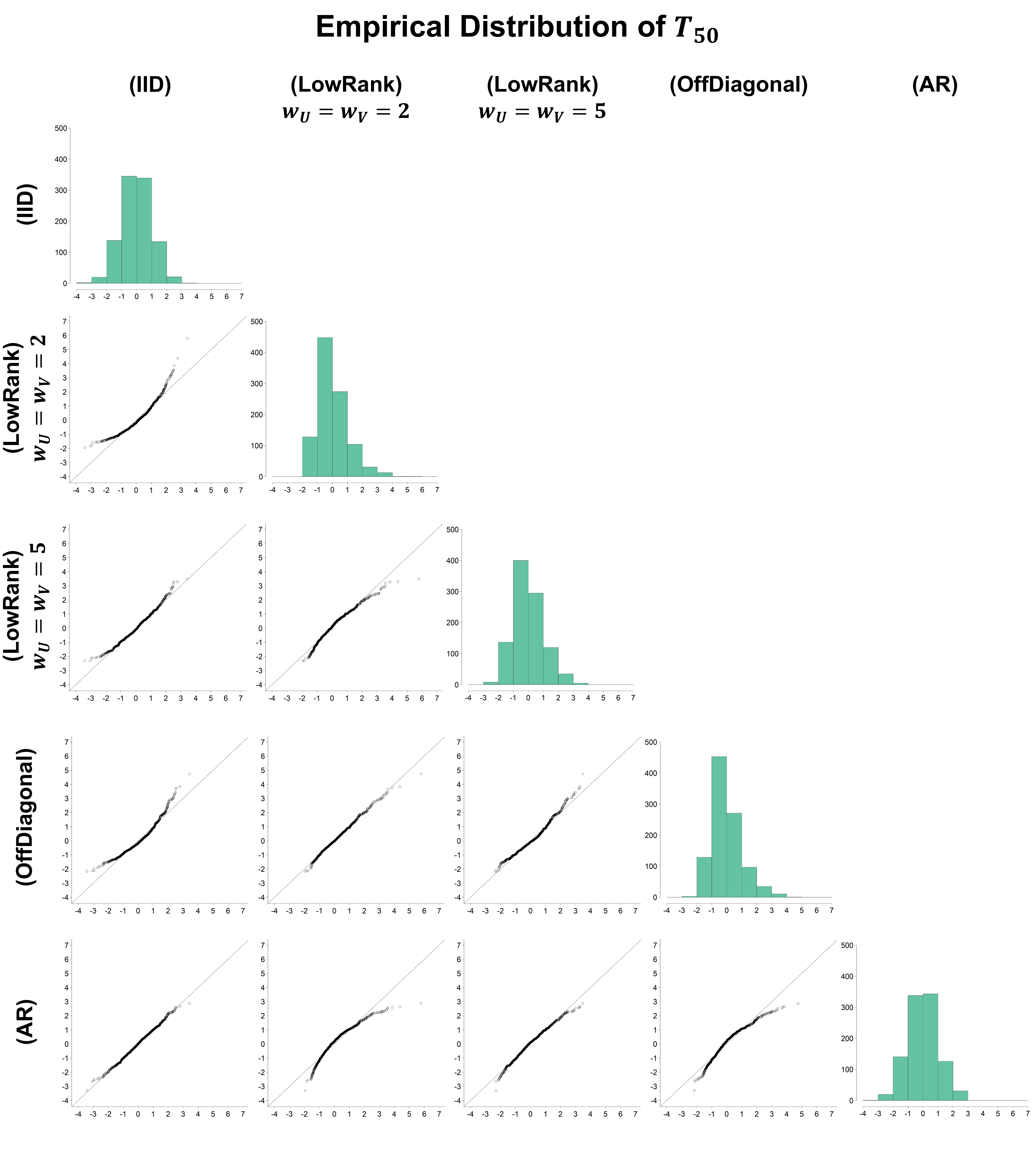}
     \caption{Empirical distribution of standardized $T_{50}$ under $H_0$ for various covariance structures in the $n=1000,p=250$ setting.}
     \label{fig:tract_distribution_50}
\end{figure}

\subsection{Sensitivity analysis of cutoff used to calculate $K$}

A key hyperparameter in RACT is the choice of $K$, which sets the maximum Ky-Fan($k$) norm which enters into $T_\text{RACT}$. We select $K$ to be the smallest $K\leq \min(n,p)$ such that the variation of $\widehat{\Sigma}$ explained by its top $K$ singular values exceeds 80\%. The choice of $K$ will affect both the power and computation time of the algorithm. As demonstrated in our simulation studies, depending on the structure of $\Sigma_1-\Sigma_2$ certain Ky-Fan($k$) norms may be more powerful, and hence a selection of $K$ which includes these norms will likely improve the power of $T_\text{RACT}$. As well, given a larger $K$ requires calculating more of the singular values at each permutation, a larger $K$ will increase the computation time of RACT.

To assess RACT's sensitivity to the choice of $K$ we vary the percentage cutoff used to select $K$ and present results showing how power, the value of $K$ chosen, and computation time (using 1000 permutations) is affected for various subsample sizes and datasets from the real data analysis section of our paper.

We see from Web Table \ref{tab:ract_K_power} that across subsample sizes and datasets the power generally appears to increase sharply from 20\% to 80\%, after which any power improvement is very modest. Web Table \ref{tab:ract_K_avg_K} shows the average $K$ chosen; we note that for subsample sizes of $50$ (i.e., each group contains 50 observations) our 80\% cutoff chooses $K$ on average to be 13.1, 15.4, and 7.9 respectively across the TCGA, SPINS FA, and SPINS MD datasets. Given the dimensions of these datasets are 72, 73, and 73, and for two subsamples of size 50 the maximum rank for the difference of these covariance matrices is also equal to 72, 73, and 73, it does appear the majority of the variation is found in a limited number of singular values. Finally Web Table \ref{tab:ract_K_time} shows the computation time of the RACT algorithm for different percentage cutoffs.

\begin{table}[H]
\centering
\caption{Power of RACT while varying the percentage cutoff used to calculate $K$ for different subsample sizes and datasets. } 
\label{tab:ract_K_power}
\small
\begin{tabular}{cccccccccc}
  \hline
Subsample Size & Dataset & 20\% & 40\% & 60\% & 80\% & 90\% & 95\% & 99\% & 100\% \\ 
  \hline
15 & TCGA & 0.157 & 0.188 & 0.218 & 0.252 & 0.265 & 0.268 & 0.268 & 0.268 \\ 
  20 & TCGA & 0.185 & 0.247 & 0.285 & 0.352 & 0.371 & 0.385 & 0.388 & 0.388 \\ 
  25 & TCGA & 0.194 & 0.297 & 0.387 & 0.467 & 0.512 & 0.517 & 0.520 & 0.520 \\ 
  30 & TCGA & 0.251 & 0.380 & 0.478 & 0.584 & 0.619 & 0.637 & 0.641 & 0.642 \\ 
  35 & TCGA & 0.292 & 0.449 & 0.567 & 0.687 & 0.730 & 0.744 & 0.747 & 0.747 \\ 
  40 & TCGA & 0.327 & 0.512 & 0.654 & 0.775 & 0.816 & 0.832 & 0.836 & 0.836 \\ 
  45 & TCGA & 0.349 & 0.549 & 0.702 & 0.826 & 0.871 & 0.883 & 0.891 & 0.892 \\ 
  50 & TCGA & 0.403 & 0.634 & 0.767 & 0.899 & 0.930 & 0.936 & 0.939 & 0.939 \\ \hline
  15 & SPINS FA & 0.193 & 0.218 & 0.310 & 0.330 & 0.333 & 0.331 & 0.332 & 0.332 \\ 
  20 & SPINS FA & 0.238 & 0.290 & 0.430 & 0.474 & 0.481 & 0.484 & 0.483 & 0.483 \\ 
  25 & SPINS FA & 0.294 & 0.348 & 0.528 & 0.596 & 0.601 & 0.603 & 0.604 & 0.604 \\ 
  30 & SPINS FA & 0.349 & 0.450 & 0.702 & 0.756 & 0.766 & 0.769 & 0.770 & 0.770 \\ 
  35 & SPINS FA & 0.413 & 0.513 & 0.784 & 0.844 & 0.845 & 0.846 & 0.845 & 0.845 \\ 
  40 & SPINS FA & 0.446 & 0.574 & 0.863 & 0.906 & 0.909 & 0.910 & 0.911 & 0.911 \\ 
  45 & SPINS FA & 0.525 & 0.675 & 0.939 & 0.968 & 0.971 & 0.970 & 0.970 & 0.970 \\ 
  50 & SPINS FA & 0.573 & 0.726 & 0.984 & 0.994 & 0.994 & 0.995 & 0.995 & 0.995 \\ \hline
  15 & SPINS MD & 0.698 & 0.700 & 0.717 & 0.772 & 0.789 & 0.790 & 0.789 & 0.789 \\ 
  20 & SPINS MD & 0.868 & 0.867 & 0.887 & 0.922 & 0.929 & 0.930 & 0.930 & 0.930 \\ 
  25 & SPINS MD & 0.935 & 0.936 & 0.956 & 0.981 & 0.981 & 0.981 & 0.982 & 0.982 \\ 
  30 & SPINS MD & 0.986 & 0.986 & 0.995 & 1.000 & 1.000 & 1.000 & 1.000 & 1.000 \\ 
  35 & SPINS MD & 0.993 & 0.993 & 0.997 & 0.999 & 1.000 & 1.000 & 1.000 & 1.000 \\ 
  40 & SPINS MD & 0.999 & 0.999 & 0.999 & 1.000 & 1.000 & 1.000 & 1.000 & 1.000 \\ 
  45 & SPINS MD & 1.000 & 1.000 & 1.000 & 1.000 & 1.000 & 1.000 & 1.000 & 1.000 \\ 
  50 & SPINS MD & 0.999 & 0.999 & 1.000 & 1.000 & 1.000 & 1.000 & 1.000 & 1.000 \\ 
   \hline
\end{tabular}\end{table}

\begin{table}[H]
\centering
\caption{Average $K$ selected by RACT using different percentage cutoffs for different subsample sizes and datasets.} 
\small
\label{tab:ract_K_avg_K}
\begin{tabular}{cccccccccc}
  \hline
Subsample Size & Dataset & 20\% & 40\% & 60\% & 80\% & 90\% & 95\% & 99\% & 100\% \\ 
  \hline
15 & TCGA & 1.0 & 2.3 & 4.5 & 8.8 & 13.2 & 17.3 & 24.3 & 29.1 \\ 
  20 & TCGA & 1.0 & 2.5 & 4.9 & 9.9 & 15.3 & 20.5 & 30.1 & 39.1 \\ 
  25 & TCGA & 1.0 & 2.6 & 5.2 & 10.8 & 16.9 & 23.0 & 35.0 & 49.1 \\ 
  30 & TCGA & 1.0 & 2.7 & 5.4 & 11.5 & 18.2 & 25.0 & 38.8 & 59.1 \\ 
  35 & TCGA & 1.0 & 2.7 & 5.6 & 12.0 & 19.2 & 26.6 & 41.8 & 69.0 \\ 
  40 & TCGA & 1.0 & 2.8 & 5.8 & 12.4 & 20.1 & 27.9 & 44.3 & 72.0 \\ 
  45 & TCGA & 1.0 & 2.8 & 5.9 & 12.8 & 20.8 & 29.0 & 46.3 & 72.0 \\ 
  50 & TCGA & 1.0 & 2.9 & 6.0 & 13.1 & 21.4 & 30.0 & 47.9 & 72.0 \\ \hline
  15 & SPINS FA & 1.0 & 1.5 & 4.0 & 9.4 & 14.7 & 19.1 & 25.8 & 29.2 \\ 
  20 & SPINS FA & 1.0 & 1.5 & 4.4 & 10.9 & 17.4 & 23.2 & 32.6 & 39.1 \\ 
  25 & SPINS FA & 1.0 & 1.6 & 4.7 & 12.0 & 19.6 & 26.5 & 38.3 & 49.1 \\ 
  30 & SPINS FA & 1.0 & 1.6 & 5.0 & 13.0 & 21.5 & 29.2 & 42.9 & 59.1 \\ 
  35 & SPINS FA & 1.0 & 1.6 & 5.1 & 13.7 & 22.8 & 31.3 & 46.5 & 69.0 \\ 
  40 & SPINS FA & 1.0 & 1.6 & 5.3 & 14.3 & 24.1 & 33.1 & 49.4 & 73.0 \\ 
  45 & SPINS FA & 1.0 & 1.6 & 5.4 & 14.9 & 25.2 & 34.6 & 51.7 & 73.0 \\ 
  50 & SPINS FA & 1.0 & 1.7 & 5.6 & 15.4 & 26.0 & 35.9 & 53.5 & 73.0 \\ \hline
  15 & SPINS MD & 1.0 & 1.1 & 2.1 & 5.9 & 10.6 & 15.4 & 23.7 & 29.5 \\ 
  20 & SPINS MD & 1.0 & 1.0 & 2.2 & 6.5 & 12.2 & 18.1 & 29.3 & 39.7 \\ 
  25 & SPINS MD & 1.0 & 1.0 & 2.3 & 6.8 & 13.3 & 20.2 & 33.8 & 49.6 \\ 
  30 & SPINS MD & 1.0 & 1.0 & 2.3 & 7.1 & 14.2 & 21.8 & 37.4 & 59.1 \\ 
  35 & SPINS MD & 1.0 & 1.0 & 2.3 & 7.4 & 14.9 & 23.2 & 40.2 & 69.0 \\ 
  40 & SPINS MD & 1.0 & 1.0 & 2.3 & 7.6 & 15.5 & 24.3 & 42.5 & 73.0 \\ 
  45 & SPINS MD & 1.0 & 1.0 & 2.3 & 7.8 & 16.0 & 25.3 & 44.4 & 73.0 \\ 
  50 & SPINS MD & 1.0 & 1.0 & 2.3 & 7.9 & 16.5 & 26.0 & 45.9 & 73.0 \\ 
   \hline
\end{tabular}
\end{table}

\begin{table}[H]
\centering
\caption{Average computation time in seconds of RACT while varying the percentage cutoff used to calculate $K$ for different subsample sizes and datasets.} 
\label{tab:ract_K_time}
\small
\begin{tabular}{cccccccccc}
  \hline
Subsample Size & Dataset & 20\% & 40\% & 60\% & 80\% & 90\% & 95\% & 99\% & 100\% \\ 
  \hline
15 & TCGA & 2.0 & 2.0 & 2.3 & 2.3 & 2.3 & 2.3 & 2.4 & 2.6 \\ 
  20 & TCGA & 2.1 & 2.1 & 2.3 & 2.5 & 2.5 & 2.5 & 2.7 & 2.8 \\ 
  25 & TCGA & 2.1 & 2.2 & 2.4 & 2.6 & 2.7 & 2.7 & 2.9 & 2.9 \\ 
  30 & TCGA & 2.2 & 2.3 & 2.5 & 2.7 & 2.8 & 2.8 & 3.0 & 3.1 \\ 
  35 & TCGA & 2.3 & 2.4 & 2.7 & 2.8 & 3.0 & 3.0 & 3.2 & 3.2 \\ 
  40 & TCGA & 2.4 & 2.7 & 2.8 & 2.9 & 3.1 & 3.1 & 3.3 & 3.5 \\ 
  45 & TCGA & 2.5 & 2.8 & 2.8 & 3.0 & 3.2 & 3.2 & 3.3 & 3.5 \\ 
  50 & TCGA & 2.5 & 2.8 & 2.9 & 3.1 & 3.3 & 3.3 & 3.4 & 3.6 \\ \hline
  15 & SPINS FA & 2.2 & 2.2 & 2.3 & 2.4 & 2.1 & 2.2 & 2.5 & 2.6 \\ 
  20 & SPINS FA & 2.3 & 2.3 & 2.4 & 2.6 & 2.7 & 2.4 & 2.7 & 2.8 \\ 
  25 & SPINS FA & 2.3 & 2.4 & 2.5 & 2.7 & 2.8 & 2.6 & 3.0 & 3.0 \\ 
  30 & SPINS FA & 2.5 & 2.5 & 2.6 & 2.8 & 2.8 & 2.8 & 3.2 & 3.2 \\ 
  35 & SPINS FA & 2.6 & 2.6 & 2.8 & 3.0 & 3.3 & 3.0 & 3.4 & 3.4 \\ 
  40 & SPINS FA & 2.7 & 2.7 & 2.8 & 3.1 & 3.1 & 3.1 & 3.5 & 3.5 \\ 
  45 & SPINS FA & 2.7 & 2.7 & 2.9 & 3.2 & 3.2 & 3.5 & 3.6 & 3.6 \\ 
  50 & SPINS FA & 2.8 & 2.8 & 3.0 & 3.3 & 3.4 & 3.6 & 3.6 & 3.7 \\ \hline
  15 & SPINS MD & 2.2 & 2.0 & 2.2 & 2.3 & 2.4 & 2.3 & 2.5 & 2.6 \\ 
  20 & SPINS MD & 2.3 & 2.1 & 2.3 & 2.4 & 2.6 & 2.6 & 2.7 & 2.8 \\ 
  25 & SPINS MD & 2.3 & 2.1 & 2.4 & 2.5 & 2.7 & 2.9 & 2.9 & 3.0 \\ 
  30 & SPINS MD & 2.5 & 2.2 & 2.5 & 2.6 & 2.8 & 3.0 & 3.1 & 3.2 \\ 
  35 & SPINS MD & 2.6 & 2.6 & 2.6 & 2.9 & 3.0 & 3.3 & 3.4 & 3.4 \\ 
  40 & SPINS MD & 2.6 & 2.4 & 2.7 & 2.9 & 3.1 & 3.5 & 3.4 & 3.5 \\ 
  45 & SPINS MD & 2.7 & 2.5 & 2.7 & 2.9 & 3.2 & 3.5 & 3.5 & 3.7 \\ 
  50 & SPINS MD & 2.8 & 2.5 & 2.8 & 3.0 & 3.3 & 3.6 & 3.6 & 3.7 \\ 
   \hline
\end{tabular}
\end{table}

\subsection{Comparison of RACT and HC}

We find in the real data analysis that RACT and HC both perform particularly well, and have roughly equal performance on the SPINS datasets. To examine the drivers behind each method's performance, in Web Figure \ref{fig:hc_ract_compare} we present the power of selected individual test statistics which make up both methods, as well as the power from both methods. We note that in implementing a permutation-based version of HC we take a minimum $p$-value across superdiagonals $q \in \{0,1,\dots, \lfloor p^{0.7} \rfloor \}$, therefore like RACT we would expect the power of HC to fall somewhere near the power of the best performing superdiagonal test statistic. We can see that for the TCGA and SPINS FA datasets HC's power is driven by superdiagonals with small values of $q$ (particularly $q=0$). On the other hand, for the SPINS MD dataset the power is spread more evenly among values of $q$. In Figure 4 of the main article the difference in covariance of the SPINS FA and SPINS MD datasets are presented without reordering. We see the structure of their difference aligns with the performance of the individual superdiagonal test statistics. For SPINS FA the covariance difference contains several several prominent diagonal blocks, driving the performance of the superdiagonal test statistics for small values of $q$. For SPINS MD, the difference is less concentrated along the main diagonal, and hence the power of the superdiagonal test statistics is more uniform across various values of $q$. For RACT, we see that the higher Ky-Fan($k$) norms contribute significantly in all cases, although in SPINS MD the improvement over lower Ky-Fan($k$) norms is modest.

\begin{figure}[H]
\centering
\includegraphics[width=\textwidth]{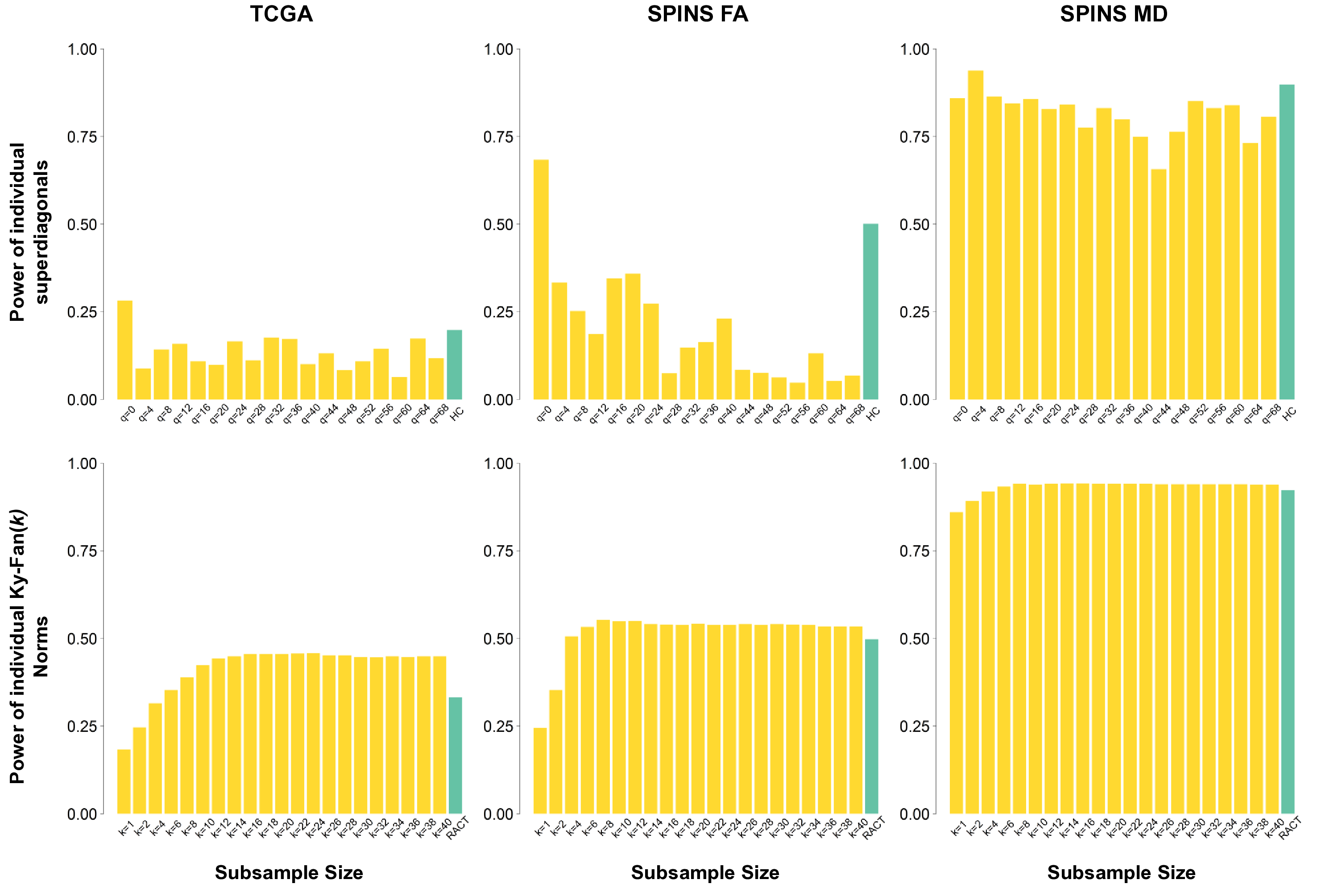}
	\caption{First row: empirical power of tests statistics based on individual superdiagionals relative to HC's overall power for a subsample size of 20. Second row: empirical power of individual Ky-Fan($k$) norms relative to RACT's individual power for a subsample size of 20. } 
 \label{fig:hc_ract_compare}
\end{figure}

\end{document}